\pdfoutput=1

\documentclass[11pt]{article}

\usepackage[final]{acl}
\usepackage{amsmath}
\usepackage{times}
\usepackage{latexsym}
\usepackage{graphicx}
\usepackage{tabularx}
\usepackage{array}

\newcolumntype{C}[1]{>{\centering\arraybackslash}m{#1}}
\usepackage{multirow}

\usepackage[T1]{fontenc}
\usepackage[utf8]{inputenc}
\usepackage{microtype}
\usepackage{inconsolata}
\usepackage{booktabs}
\usepackage{hyperref}

\title{\includegraphics[width=0.06\textwidth]{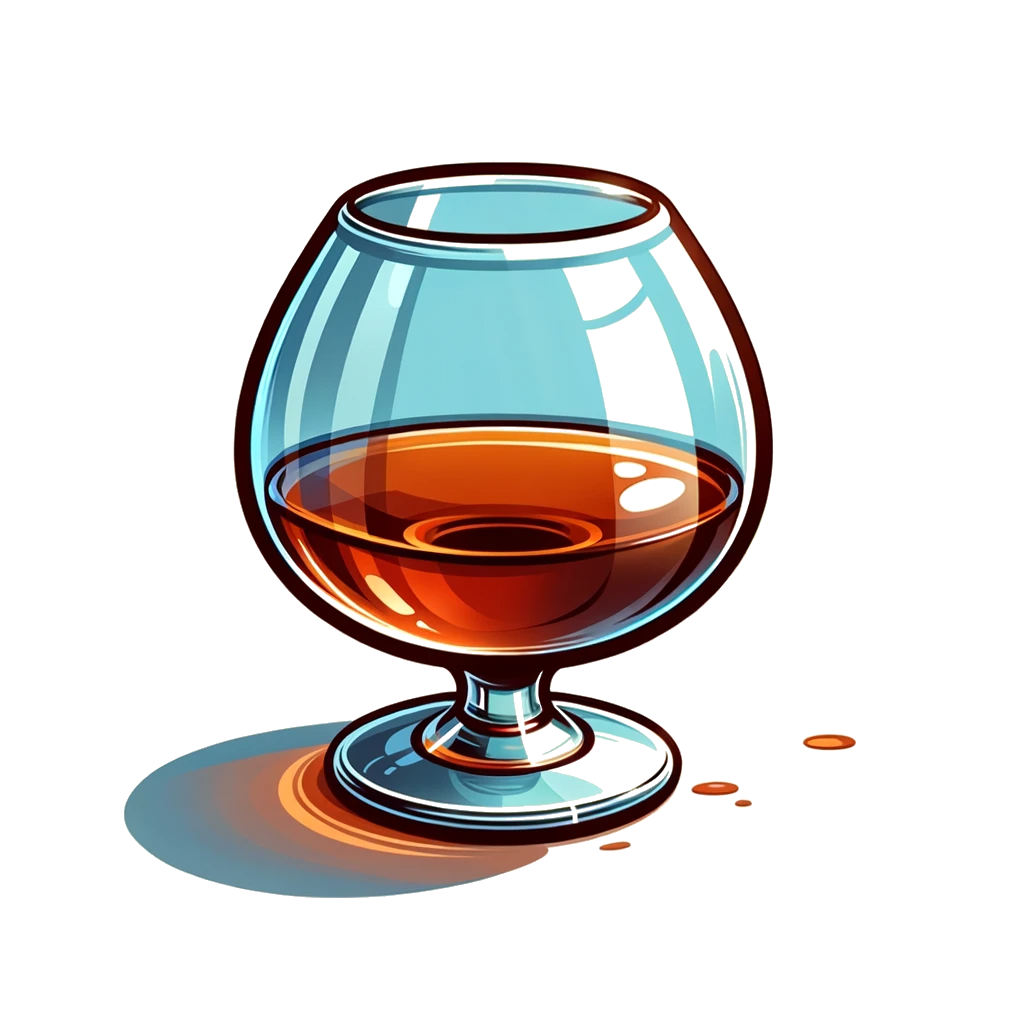} BIRCO: A Benchmark of Information Retrieval Tasks with Complex Objectives}

\author{Xiaoyue Wang \thanks{Equal contribution} \quad
        Jianyou Wang\footnotemark[1] \quad
        Weili Cao\footnotemark[1] \quad
        Kaicheng Wang\footnotemark[1] \\
        \textbf{Ramamohan Paturi \thanks{Equal Senior Authors} \quad
        Leon Bergen\footnotemark[2]} \\
        \\
        Laboratory for Emerging Intelligence\\
        University of California, San Diego \\
        \texttt{\{xiw027, jiw101, w2cao, kaw036, rpaturi, lbergen\}@ucsd.edu}
}

\begin{document}
\maketitle
\begin{abstract}
We present the \textbf{B}enchmark of \textbf{I}nformation \textbf{R}etrieval (IR) tasks with \textbf{C}omplex \textbf{O}bjectives (BIRCO). BIRCO evaluates the ability of IR systems to retrieve documents given multi-faceted user objectives. The benchmark's complexity and compact size make it suitable for evaluating large language model (LLM)-based information retrieval systems. We present a modular framework for investigating factors that may influence LLM performance on retrieval tasks, and identify a simple baseline model which matches or outperforms existing approaches and more complex alternatives. No approach achieves satisfactory performance on all benchmark tasks, suggesting that stronger models and new retrieval protocols are necessary to address complex user needs. \footnote{\href{https://github.com/BIRCO-benchmark/BIRCO}{https://github.com/BIRCO-benchmark/BIRCO}}

\end{abstract}

\section{Introduction}

Information retrieval (IR) tasks have traditionally been centered around matching queries with semantically similar passages. However, user objectives may go significantly beyond retrieving based on similarity. As a motivating example, consider a user who wants to find papers that refute a particular scientific claim. This would not be well-captured by similarity-driven search, which would also retrieve papers that support the claim. In addition, the user may have multiple objectives in their search. They may be searching for papers that measure the response of a drug in a specific population, using a certain set of measurements.

We propose the BIRCO benchmark for evaluating the performance of IR systems on tasks with \emph{complex objectives}.
We curate 5 open-source datasets (DORIS-MAE \cite{dorismae}, ArguAna \cite{arguana}, WhatsThatBook \cite{wtb}, Clinical-Trial \cite{clinical-trial}, and RELIC \cite{relic}), which contain paragraph-length queries with multi-faceted task objectives. This represents a challenging test bed for methods that aim to address complex user search needs.

IR systems fall into three primary categories: pre-trained embedding models, language models that have been fine-tuned for IR tasks, and task-agnostic models based on Large Language Models (LLMs) like GPT4.

BIRCO provides a benchmark for LLM-based retrieval systems. The scale and capabilities of these models present a unique set of challenges for IR benchmarks. Due to their massive pretraining, LLMs may be able to answer many user queries without examining the corresponding documents. This decreases the validity of these tasks for measuring LLM retrieval performance on novel queries and document sets. In addition, IR tasks may contain thousands of documents per query, making it cost-prohibitive to evaluate LLM systems.

BIRCO is designed to address both of these challenges. The complexity of the search objectives, combined with aggressive filtering, makes it impossible for LLMs such as GPT4 to answer BIRCO queries without examining the corpus documents. BIRCO also remains challenging for LLMs despite having only 50-100 documents per query, making it low-cost to evaluate LLM performance.

\section{Related Work}

\begin{figure*}[ht]
    \centering
        \includegraphics[width=\linewidth]{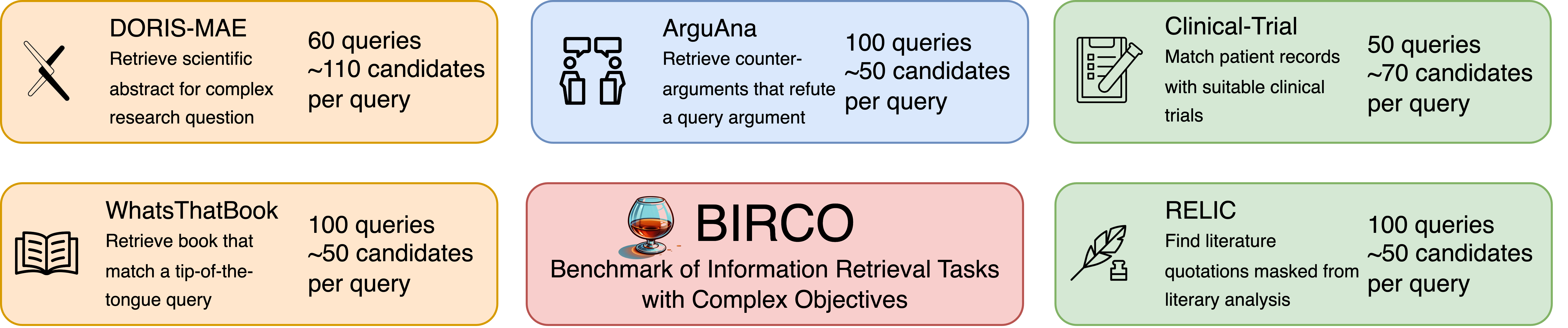}
    \caption{BIRCO contains 5 IR tasks with complex objectives}
    \label{fig: query classification}
\end{figure*}

\textbf{IR Benchmarks}\\
IR benchmarks such as MS MARCO \cite{msmarco}, NQ \cite{nq}, LOTTE \cite{colbertv2}, BEIR \cite{beir}, MIRACL \cite{miracl}, MTEB \cite{mteb}, and BERRI \cite{asai-etal-2023-task} consist mostly of sentence-level queries, and their task objectives, while varying to some degree, focus on finding semantically similar passages, with one exception: the ArguAna dataset \cite{arguana}, which is a counterargument retrieval task.  \\
\textbf{Complex Query IR Tasks}\\
Several recent datasets (DORIS-MAE, WhatsThatBook, Clinical-Trial, RELIC) pose more complex retrieval tasks \cite{dorismae, wtb, clinical-trial, relic}. In these datasets, the queries are paragraph-length, and passages should match the queries along multiple dimensions. See Figure \ref{fig: query classification}. \\
\textbf{Specialized retrieval models}\\
Pretrained \cite{ada, e5, simcse} and fine-tuned \cite{rankcse, diffcse, promptagator, HyDE, exaranker, controlretriever} embedding models have formed the core of most IR systems due to their speed and simplicity. More recently, there have been methods for fine-tuning language models for ranking and retrieval, including monoT5 \cite{monot5} and RankLLaMA \cite{rankllama}. TART \cite{asai-etal-2023-task} and INSTRUCTOR \cite{su-etal-2023-one} are trained to follow task-specific instructions during retrieval.\\
\textbf{LLM-based IR systems}\\
Recent research has shown that LLMs can be effectively used for the re-ranking stage of IR. \citet{qg,helm} compute a relevance score with output logits, \citet{prp} use pairwise comparison among passages with open-source LLMs, \citet{rankgpt, lrl} use list-wise comparisons, and \citet{rg} have the LLM assign a 4-way label to each query/passage pair. These methods have primarily been evaluated on sentence-level queries. 
\textbf{Benchmarks for LLMs}\\
Recent LLMs are evaluated across various datasets for specific skills. Big-Bench Hard \cite{Big_Bench_hard} assesses model’s abilities in handling challenging tasks that require multi-step reasoning. DROP \cite{DROP} evaluates capabilities in reading comprehension, while HellaSwag \cite{Hellaswag} examines commonsense reasoning skills. Meanwhile, MMLU \cite{MMLU} employs multiple-choice problems to analyze LLMs' problem-solving skills across a wide range of subjects. To the best of our knowledge, no IR benchmark designed to evaluate LLM-based IR system is released or documented.

\section{A Benchmark for LLM-based IR Systems}

\begin{table*}[htbp]
\centering
\scriptsize
\begin{tabular}{p{1.5cm}p{1.0cm}p{3.0cm}p{1.0cm}C{0.8cm}|C{0.6cm}C{0.6cm}C{0.9cm}C{0.7cm}C{0.7cm}C{0.9cm}}
\toprule

\multirow{2}{*}[-1.75em]{\textbf{Dataset}} & \multirow{2}{*}[-1.75em]{\textbf{Domain}}& \multirow{2}{*}[-1.75em]{\textbf{Task Objective}} & \multirow{2}{*}[-1.75em]{\textbf{Relevancy}} & \multicolumn{1}{c}{\textbf{Dev}} & \multicolumn{6}{c}{\textbf{Test}} \\ 
\cmidrule(lr){5-5}\cmidrule(lr){6-11}
& & & & Num. of Triplets & Num. of Queries & Num. of Docs & Relevant Doc. per Query & Query Length & Doc Length & Lexical Overlap\\
 
\midrule
\multicolumn{11}{c}{\textbf{BIRCO}}\\
\midrule
DORIS-MAE & AI/ML & Retrieve relevant abstracts for the research question. & Multi-level & 3k & 60 & 110.6 & 18.2 & 166.7 & 199.5 & 0.147\\
ArguAna & Debate &  Retrieve the most effective argument that counters the query. & Binary & 19k & 100 & 50.0 & 1 & 189.1 & 185.5 & 0.139\\
WhatsThatBook & Literature & Find the book from user's ambiguous descriptions.
 & Binary& 3k & 100 & 50.4 & 1 & 152.2 & 189.3 & 0.049\\

Clinical-Trial & BioMed & Match patient records with suitable clinical trials. & 3-level & 1k &  50 & 68.4 & 19.7 & 78.5 & 175.6& 0.050\\
RELIC & Literature & Recover the masked quotation in a literary analysis. & Binary & 68k & 100 & 50.6 & 1 & 169.1 & 85.4 & 0.027\\

\midrule 
\textbf{BIRCO Avg.} & - & - & - & 20k & 51.3 & 66.0 & 8.4 & 168.3 & 185.1 & 0.083  \\
\midrule

\textbf{BEIR Avg.} & - & - & - & - & 3710.9 & 2417k & 43.5 & 9.8 & 124.9 & 0.384\\

\bottomrule
\end{tabular}
\caption{BIRCO Dataset Statistics and Comparison with BEIR Statistics. Relevant Doc. per Query means the number of relevant passages per query. Lexical overlap is calculated as the number of overlapping monograms between query and its relevant passage(s) without counting stopwords. Num of Triplets is the number of triplets extracted for the dev set. See Appendix \ref{sec: appendix_dev_set} for details of triplet extraction.}
\label{tab: birco-beir-comparison}
\end{table*}

\begin{table*}[ht]
\centering
\footnotesize
\setlength{\tabcolsep}{2.0pt}
\begin{tabular}{@{}cc|cccc|cccc|c@{}}
\toprule
&\multicolumn{1}{c|}{MS MARCO} & \multicolumn{1}{c}{BIRCO} & \phantom{abc} && \multicolumn{1}{c|}{NQ} &  \multicolumn{1}{c}{BIRCO} & \phantom{abc} && \multicolumn{1}{c}{BEIR} & \multicolumn{1}{c}{BIRCO}\\
Models & MRR@10 & MRR@10 && Models & R@20 & R@20 && Models & nDCG@10 & nDCG@10 \\
\cmidrule{1-3} \cmidrule{5-7}\cmidrule{9-11}

Max possible & 100.0 & 100.0 && Max possible & 100.0 & 94.3 && Max possible & 100.0 & 100.0\\
ANCE\textsubscript{FirstP} & 33.0 & 20.0 && ANCE\textsubscript{FirstP} & 81.9 & 49.6 && E5-L-v2 & 50.0 & 38.5\\
SimLM & 41.1 & 18.1 && SimLM & 85.2 & 44.6 && RankLLaMA & 56.6 & 47.4\\
SPLADE-v2 & 36.8 & 17.7 &&BM25 & 59.1 & 33.5  &&TART & 44.8 & 39.5\\

\bottomrule
\end{tabular}
\caption{Model performance on BIRCO and other IR datasets. Models and metrics are chosen based on availability of data in the published literature. Max possible means the maximum possible value for a metric on a dataset.}
\label{tab: nq_msmarco_beirs}

\end{table*}

\begin{figure*}[ht]
    \centering
        \includegraphics[width=\linewidth]{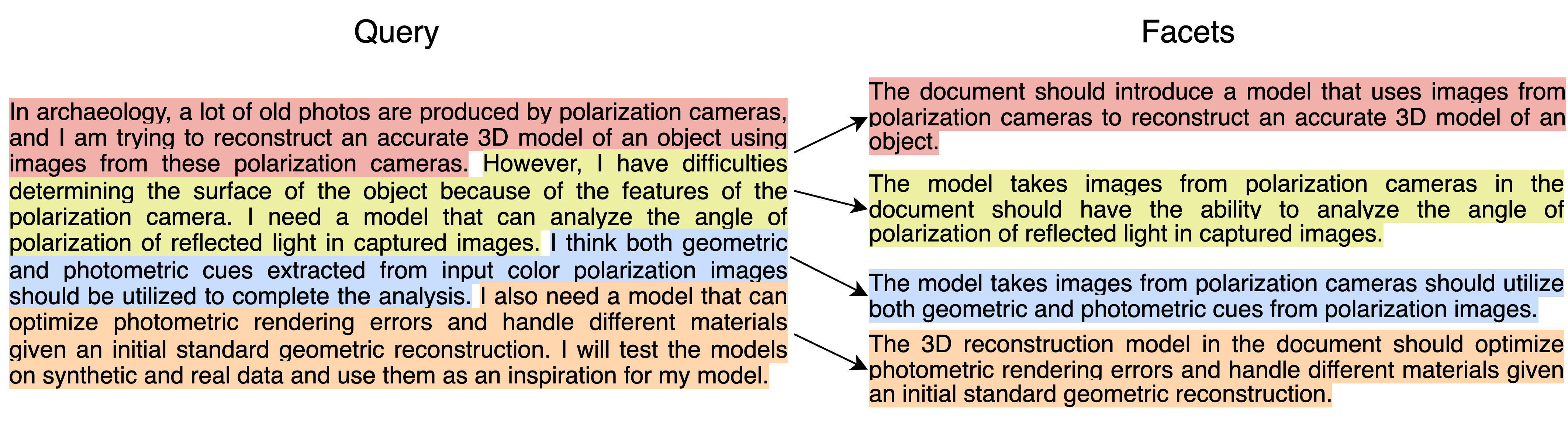}
    \caption{An example query from DORIS-MAE \cite{dorismae} with multiple facets.}
    \label{fig: facet_example}
\end{figure*}

BIRCO is designed to evaluate LLM-based IR Systems. It has several features that distinguish it from traditional IR benchmarks: longer query lengths, multiple facets within each query, complex and varied task objectives (beyond semantic similarity), minimal contamination with respect to external knowledge, increased task difficulty, and a smaller candidate pool to reduce the cost of LLM-based evaluations.

\subsection{Dataset Descriptions}
See Figure \ref{fig: query classification} for an overview of the datasets.\\
\textbf{DORIS-MAE} \cite{dorismae}\\
60 queries that are complex research questions from computer scientists. The query communicates specific requirements from research papers. Candidate pools have approximately 110 documents.  \\
\textbf{ArguAna} \cite{arguana}\\
100 queries, each with a candidate pool of around 50 passages. Queries and passages are both complex one-paragraph arguments about current affairs. The objective is to find matching counterarguments. \\
\textbf{WhatsThatBook} \cite{wtb}\\
100 queries, with each query describing a book in an ambiguous manner. Each query has a pool of 50 passages, which are book descriptions.\\
\textbf{Clinical-Trial} \cite{clinical-trial}\\
100 queries that are paragraph-length patient case reports. Each query has a candidate pool comprising 30-110 passages that are paragraph-length descriptions of clinical trials. The objective is to find the most suitable clinical trial for a patient. \\
\textbf{RELIC} \cite{relic}\\
100 queries which are excerpts from scholars analyzing classic English-language literature. Passages are sentences from a novel that have been extracted from the queries. The objective is to match a literary analysis with its missing quotations.
See Appendix \ref{sec: appendix_example} for examples of each dataset.

\subsection{BIRCO Evaluation Methodology}
BIRCO is constructed to allow for statistically valid evaluation of model performance. Four of the five benchmark datasets do not have previously defined development/test set splits. We, therefore, define splits for these datasets, ensuring that there is no overlap between queries or passages across the splits. Since we prioritize the construction of test set, and to ensure no overlap between passages in test set and development set, we remove all passages in development set candidate pools that appear in the test set. Details of development sets can be found in Appendix \ref{sec: appendix_dev_set}.

See Table \ref{tab: birco-beir-comparison} for general statistics comparing BIRCO with BEIR benchmark datasets.

\subsection{Data Decontamination}
Due to the massive scale of LLM pretraining, there is a risk of data contamination when evaluating LLM-based systems on IR benchmarks. This will lead to inflated estimates of their performance on novel retrieval tasks. We first measure the degree of contamination on existing tasks on BEIR, and then describe the decontamination procedure used for BIRCO.

\subsubsection{Measuring contamination}
We measured contamination in the prior benchmark BEIR by evaluating whether an LLM is able to correctly answer a query without access to the pool of candidate documents. If this occurs, then this transforms the retrieval task, and means that it no longer measures the model's ability to compare queries and documents.

\begin{samepage}

GPT4 was prompted with a query from BEIR (see prompts in Appendix \ref{ssec: appendix_prompts_for_decontamination}), and its response was then compared with the ground-truth answers provided in the dataset.\footnote{In some cases, the dataset provides multiple ground truth answers. When this occurred, the LLM-generated answer was compared with the ground truth answer with the highest similarity, as measured by the ada-002 embedding model.} 
\end{samepage}
The responses from GPT4 were binned into three categories: GPT4 agreed with the ground-truth answer (Category 0); GPT4 and the ground-truth answer disagreed, but GPT4's answer was also correct (Category 1); GPT4 and the ground-truth answer disagreed, and the ground-truth answer was better than GPT4's (Category 2).

The experiment was performed on 10 datasets from BEIR: MS MARCO \cite{msmarco}, NQ \cite{nq}, HotpotQA \cite{hotpotqa}, FiQA \cite{fiqa}, DBPedia \cite{dbp}, TREC-COVID \cite{trec-covid}, FEVER \cite{fever}, Climate-FEVER \cite{climate-fever}, Touché-2020 \cite{touche}, and SciFact \cite{scifact}. For each dataset, 100 queries were randomly sampled\footnote{TREC-COVID and Touché only have 50 queries, so all queries were used for this dataset.}, and two human annotators performed the labeling.

In order to evaluate the reliability of the labeling process, we measured inter-annotator agreement on a separate set of 100 queries from these 10 datasets. Labels from the annotators exactly matched 91\% of the time, had a Spearman correlation of 0.81, and had a macro F1-score of 0.76, indicating strong agreement between the annotators.

Table \ref{tab: beir-contamination} shows that GPT4 is able to answer a large fraction of queries directly, without access to candidate documents. This suggests that data contamination is a challenge for existing benchmarks aiming to evaluate LLM performance on information retrieval tasks.

\subsubsection{Reducing contamination}

In order to address the potential for similar contamination in BIRCO, we designed a procedure to remove all queries that the LLM could potentially answer without access to candidate passages. GPT4 was prompted with a query as well as task-specific instructions. For example, for Clinical-Trial, the model was given a description of a patient, and was instructed to generate a clinical trial that would be appropriate for the patient. A query was filtered when the GPT-generated passage was similar to the ground-truth passage. See Appendix \ref{sec: appendix_decontamination} for the filtering criteria for each task.

The filtering process removed 7\% of queries from Clinical-Trial; 8\% from RELIC; 56\% from WhatsThatBook; and none from DORIS-MAE and ArguAna. The WhatsThatBook task is retrieving literary novels, a topic that GPT4 has extensive knowledge of, leading to a high filtering rate. In contrast, the DORIS-MAE task is to retrieve specific computer science articles, which appears to be too fine-grained for GPT4 to answer without a candidate pool. Similarly, for ArguAna, the task is to retrieve counterarguments given an argument. These counterarguments typically include detailed factual information which GPT4 cannot generate given the argument alone.

\begin{table}[ht]
\centering
\begin{minipage}{0.23\textwidth}
\centering
\scriptsize
\setlength{\tabcolsep}{3.5pt}
\begin{tabular}{l ccc}
\toprule
\textbf{Dataset/Level} & \textbf{C0} & \textbf{C1} & \textbf{C2} \\
\midrule


MS MARCO & 76\% & 18\% & 6\% \\
TREC-COVID & 94\% & 4\% & 2\% \\
NQ & 89\% & 7\% & 4\% \\
HotpotQA & 60\% & 20\% & 20\% \\
FiQA & 73\% & 19\% & 8\% \\
\bottomrule
\end{tabular}
\end{minipage}%
\hfill
\begin{minipage}{0.23\textwidth}
\centering
\scriptsize
\setlength{\tabcolsep}{3.5pt}
\begin{tabular}{l ccc}
\toprule
\textbf{Dataset/Level} & \textbf{C0} & \textbf{C1} & \textbf{C2} \\
\midrule

DBPedia & 78\% & 12\% & 10\% \\
FEVER & 99\% & 0\% & 1\% \\
Climate-FEVER & 81\% & 18\% & 1\% \\
Touché-2020 & 98\% & 2\% & 0\% \\
SciFact & 71\% & 25\% & 4\% \\
\bottomrule
\end{tabular}
\end{minipage}
\caption{Data contamination in BEIR. GPT4 answers the query without access to any candidate passages. C0,1,2 corresponds to Category 0,1,2. C0 means contamination. Numbers are rounded to the nearest percentile.}
\label{tab: beir-contamination}
\end{table}

\subsection{Query Complexity}

\begin{table}[ht]
\centering
\setlength{\tabcolsep}{1.5pt}
\begin{tabular}{l ccc}
\toprule
\textbf{Dataset} & \textbf{Average \# Facets} & \textbf{Max} & \textbf{Min} \\ 
\midrule
DORIS-MAE & 5.6 & 9 & 3 \\ 
ArguAna & 5.8 & 11 & 2 \\ 
WhatsThatBook & 11.3 & 25 & 5 \\ 
Clinical-trial & 7.4 & 30 & 1 \\ 
RELIC & 2.0 & 2 & 1 \\ 
\bottomrule
\end{tabular}
\caption{BIRCO queries contain multiple facets. Facets represent semantic components in DORIS-MAE, arguments in ArguAna, book details in WhatsThatBook, patient symptoms in Clinical-Trial, and contrasts in the preceding and subsequent contexts in RELIC.}
\label{tab: birco-multifacetedness}
\end{table}

Table \ref{tab: birco-beir-comparison} shows that queries in BIRCO are substantially longer than those in BEIR. In addition to their length, the queries in BIRCO are multi-faceted, meaning that relevant passages must satisfy multiple requirements given by the query. We quantify the number of facets per query by prompting GPT4 to break the queries into their distinct components. See Figure \ref{fig: facet_example} for an illustration of breaking a DORIS-MAE query into 4 facets. 
Appendix \ref{sec: appendix_procedure to calculate facets from queries} provides more details about the prompting procedure. 

Table \ref{tab: birco-multifacetedness} shows the results of the facet analysis. The datasets in BIRCO have between 2 and 11 facets per query on average. IR systems must, therefore, assess documents along multiple dimensions when determining relevance.

\subsection{Task Complexity}

We use several approaches to quantify the difficulty of BIRCO. Table \ref{tab: birco-beir-comparison} shows that there is low lexical overlap between queries and relevant documents in BIRCO, making it difficult for lexical matching heuristics to succeed on the benchmark.

IR systems also achieve lower performance on BIRCO compared to other IR benchmarks, as shown in Table \ref{tab: nq_msmarco_beirs}. It is difficult to directly compare metrics across datasets, as metrics can be influenced by the number of relevant passages per query as well as the relevance scale (e.g. binary, multi-level). Table \ref{tab: nq_msmarco_beirs} uses metrics MRR@10, R@20 and nDCG@10, all of which can reach high percentages on their respective datasets. See Appendix \ref{sec: appendix_analysis of the comparison between birco and other datasets}.

\subsection{Domain Diversity}

BIRCO consists of datasets from diverse domains, including debate, computer science, biomedicine, and literature. To evaluate this diversity in terms of word overlap between each pair of datasets, we use pairwise weighted Jaccard similarity \cite{jaccard-similarity}. Figure \ref{fig: heatmap} shows low word overlap between these datasets. The most similar datasets are RELIC and WhatsThatBook, which focus on classic English literature and contemporary novels, respectively. 

\subsection{Hard Negatives in BIRCO}
\label{sec:hard-negatives}

\begin{figure*}[ht]
    \centering
    \begin{minipage}{0.49\textwidth}
        \includegraphics[width=\linewidth]{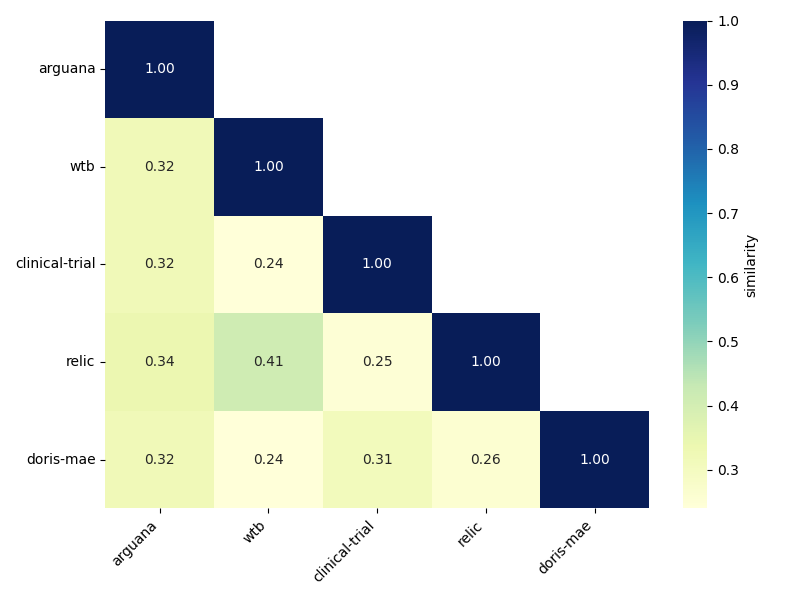}
        \caption{Jaccard similarity between pairs of datasets from BIRCO}
        \label{fig: heatmap}
    \end{minipage}\hfill
    \begin{minipage}{0.49\textwidth}
        \includegraphics[width=\linewidth]{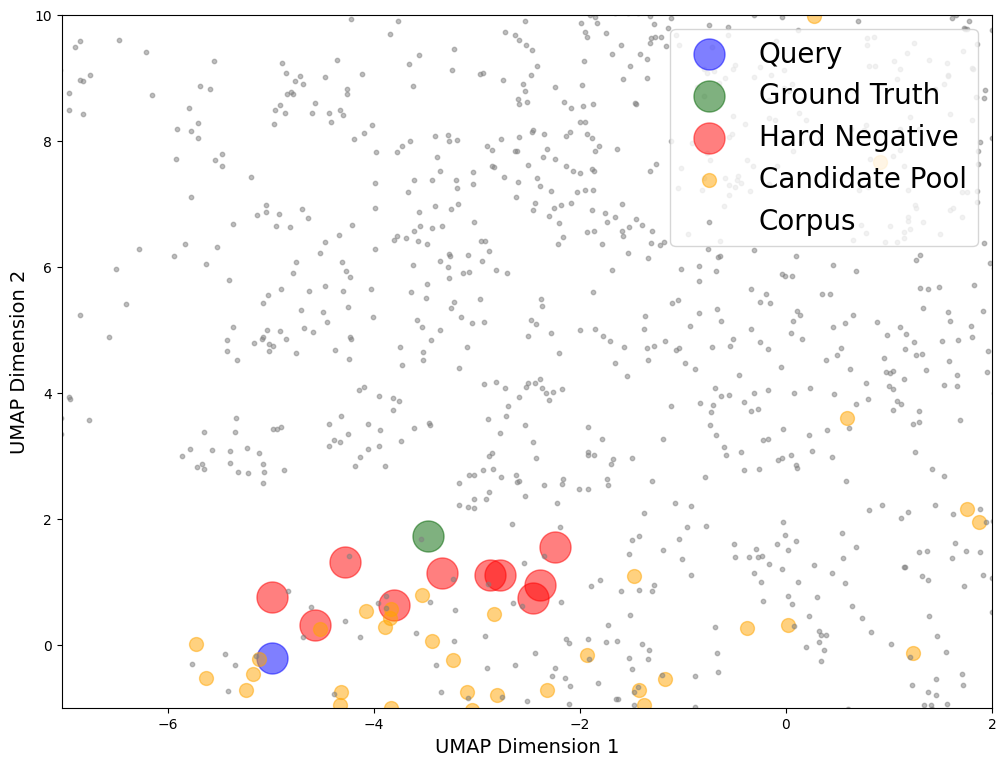}
        \caption{Passage embeddings for a randomly selected query from ArguAna. \cite{arguana}}
        \label{fig: umap}
    \end{minipage}
\end{figure*}

Due to their construction, two of the datasets in BIRCO contain hard negatives, defined as non-relevant candidate documents which share many factors in common with ground truth answers. In ArguAna, arguments and counterarguments are paragraphs which have been sampled from longer debates. Hard negatives for ArguAna are defined as candidate passages which have been extracted from the same debate as the ground truth counterargument. In Clinical-Trial, hard negatives are non-relevant clinical trials which also match the patient's disease. 22\% of the passages from ArguAna are hard negatives, and 24\% percent from Clinical-Trial.

To illustrate the effect of hard negatives, we use UMAP \cite{UMAP} to visualize the ada-002 embeddings of queries and their candidate pools. Figure \ref{fig: umap} shows a randomly sampled query from ArguAna, where its ground-truth passage is surrounded by hard negative passages.

\subsection{Candidate Pool Construction}

It is prohibitively expensive to evaluate LLMs on the entire set of passages for each query. We therefore define candidate pools for each query, restricting LLM search to these smaller pools. This is standard for many other IR tasks, where it is known as the passage re-ranking stage.

We construct candidate pools for each query using the lexicon-matching algorithm BM25 \cite{bm25_2, bm25} and state-of-the-art text embedding model ada-002 \cite{ada}. Each query has a candidate pool of 50 passages. (The ground-truth passage is inserted when necessary.) As shown in Appendix \ref{sec: appendix_the effects of candidate pool construction} and Table \ref{tab: test_set_whole_set_comparison}, the difficulty is comparable to the original datasets for four of five tasks, and remains challenging for the fifth. 

\section{A Framework for LLM-based Retrieval}
\label{sec: llm-framework}

This section presents a framework for using LLMs for information retrieval tasks. The framework is modular, allowing us to investigate the effect of different factors on model performance. See an illustration of the framework in Appendix \ref{sec: appendix_framework_illustration}.

\subsection{Ranking vs. Scoring}

LLMs can find relevant passages by either ranking them comparatively or by directly scoring them. Rank\textsubscript{\scriptsize{GPT4}} \cite{rankgpt} puts a list of passages in the LLM prompt, and iteratively filters for the most relevant passages using a sliding window approach. LLMs can also score passages one at a time. For this direct scoring approach, the LLM is prompted to generate a numerical score from 0-10.

When passages are directly scored by the LLM, there can be ties in which several passages receive the same score. By default, we use E5-Large-v2, an embedding model, for tie-breaking. 

\subsection{Chain-of-Thought Reasoning}
To investigate the role of natural language reasoning in complex query retrieval, we compare two approaches to scoring a query/passage pair. As shown in Figure \ref{fig: SER_DeAGO_Framework}, the first approach (shown in the middle of Figure \ref{fig: SER_DeAGO_Framework}, Reason+O) is to generate a set of decision criteria for judging whether a query is relevant to a passage. The LLM is instructed to follow its own decision criteria step-by-step before producing a final score. An example set of decision criteria can be seen in Figure \ref{fig: example_decision_criteria}.
Another approach (top of Figure \ref{fig: SER_DeAGO_Framework}, Score+O) is to directly produce a score given the query and passage, without any reasoning. The detailed prompt structure for these two methods is recorded in Appendix \ref{ssec: appendix_prompts_for_scoreO} and \ref{ssec: appendix_prompts_for_reasonO}.

\subsection{Task Decomposition}
With task decomposition, LLM-generated decision criteria are used to define substasks which are independently solved by the LLM. The final score is computed by averaging scores from the subtasks. This strategy, denoted as Subtask+O, aims to reduce the complexity of evaluating whether a passage is relevant to a query.

\subsection{Task Objective Awareness}
BIRCO tasks vary in their objectives, and these objectives can be described in LLM prompts. Any model that uses a prompt containing the task description is suffixed with "O". Alternatively, as a simpler baseline, LLMs can be prompted to retrieve semantically similar passages without knowing the task objective. 

All prompts that describe task objectives are based on the description of these datasets in their original papers, and tested on BIRCO's development sets. Prompts for task objectives in Appendix \ref{sec: appendix_task objective for birco}. 

\section{LLM Benchmark Results}

\begin{table*}
\footnotesize
\setlength{\tabcolsep}{1.5pt}
\centering
\begin{tabular}{l cc | cc | cc | cc | cc | C{1.5cm}}
\toprule
\textbf{\footnotesize{Model}} & \multicolumn{2}{c}{\textbf{\footnotesize{DORIS-MAE}}} & \multicolumn{2}{c}{\textbf{\footnotesize{ArguAna}}} & \multicolumn{2}{c}{\textbf{\footnotesize{WhatsThatBook}}} & \multicolumn{2}{c}{\textbf{\footnotesize{Clinical-Trial}}} & \multicolumn{2}{c}{\textbf{\footnotesize{RELIC}}}  \\
& \scriptsize{nDCG@10} & \scriptsize{Recall@5} & \scriptsize{nDCG@10} & \scriptsize{Recall@5} & \scriptsize{nDCG@10} & \scriptsize{Recall@5} & \scriptsize{nDCG@10} & \scriptsize{Recall@5} & \scriptsize{nDCG@10} & \scriptsize{Recall@5} & \scriptsize{Avg. nDCG@10} \\
\midrule
Maximum Possible & 100.0 & 45.6 & 100.0 & 100.0 & 100.0 & 100.0 & 100.0 & 43.0 & 100.0 & 100.0 & 100.0 \\
Random & 49.9 & 4.5 & 9.1  & 9.9  & 8.9 & 9.8 & 22.4  & 8.4  & 8.9  & 10.0 & 19.8 \\
\midrule
\multicolumn{12}{c}{\textbf{\footnotesize{Embedding Models}}} \\ 
E5-L-v2    &  72.0  &  14.6  &  44.9  &  62.0  &  36.6  &  39.9  &  30.4  &  11.3  &  10.9  &  14.9  &  39.0 \\ 
SIMCSE    &  70.8  &  14.4  &  40.3  &  51.9  &  31.7  &  37.0  &  27.1  &  10.7  &  12.7  &  15.9  &  36.5 \\ 
Promptagator  & 76.0 & 15.6 & 60.9 & \textbf{73.7} & 58.4 & 62.9 & NA & NA & NA & NA & NA\\
\midrule
\multicolumn{12}{c}{\textbf{\footnotesize{Finetuned Encoder-Decoder Models}}} \\ 
TART    &  64.9  &  10.2  &  49.3  &  66.7  &  42.7  &  46.1  &  32.9  &  11.1  &  9.1  &  7.0  &  39.8 \\ 
TART+O    &  49.6  &  4.9  &  23.6  &  29.9  &  17.9  &  20.7  &  24.1  &  7.5  &  13.2  &  17.1  &  25.7 \\ 
MonoT5    &  66.9  &  12.4  &  46.8  &  52.7  &  50.8  &  59.0  &  34.3  &  \textbf{15.6}  &  11.9  &  10.0  &  42.1 \\ 
\midrule
\multicolumn{12}{c}{\textbf{\footnotesize{Finetuned Decoder-Only Model}}} \\ 
RankLLaMA    &  75.1  &  \textbf{18.4}  &  54.1  &  72.6  &  63.7  &  69.8  &  30.7  &  10.9  &  14.9  &  18.0  &  47.7 \\ 
\midrule
\multicolumn{12}{c}{\textbf{\footnotesize{Comparison-based LLM IR Systems}}} \\ 
Rank\textsubscript{\scriptsize{StripedHyena}}    &  71.7  &  14.6  &  39.8  &  58.0  &  33.6  &  39.8  &  29.5  &  10.3  &  10.9  &  13.0  &  37.1 \\ 
Rank+O\textsubscript{\scriptsize{StripedHyena}}    &  70.6  &  15.5  &  38.9  &  54.8  &  32.8  &  39.9  &  29.8  &  10.8  &  9.2  &  11.9  &  36.3 \\ 
\\ 
Rank\textsubscript{\scriptsize{Llama2-7b}}    &  71.0  &  13.2  &  40.5  &  58.9  &  25.7  &  35.9  &  31.2  &  11.5  &  11.1  &  11.9  &  35.9 \\ 
Rank+O\textsubscript{\scriptsize{Llama2-7b}}    &  70.8  &  14.1  &  39.4  &  50.9  &  26.5  &  37.8  &  28.5  &  9.6  &  10.8  &  14.0  &  35.2 \\ 
\\ 
Rank\textsubscript{\scriptsize{Llama2-13b}}    &  71.2  &  13.0  &  41.1  &  57.0  &  27.1  &  37.8  &  32.9  &  13.8  &  9.9  &  14.9  &  36.4 \\ 
Rank+O\textsubscript{\scriptsize{Llama2-13b}}    &  70.5  &  14.2  &  37.8  &  47.8  &  28.0  &  38.9  &  30.0  &  10.9  &  9.2  &  12.9  &  35.1 \\ 
\\ 
Rank\textsubscript{\scriptsize{Llama2-70b}}    &  70.7  &  15.1  &  41.3  &  55.8  &  30.2  &  37.8  &  29.0  &  9.8  &  10.6  &  12.9  &  36.4 \\ 
Rank+O\textsubscript{\scriptsize{Llama2-70b}}    &  70.4  &  14.2  &  40.2  &  55.9  &  25.6  &  38.9  &  30.0  &  11.4  &  9.1  &  10.9  &  35.1 \\ 
\\ 
Rank\textsubscript{\scriptsize{GPT4}}    &  76.2   &  \textbf{17.5}   &  27.3   &  17.0   &  \textbf{80.1}   &  \textbf{86.9}  &  \textbf{40.3}   &  \textbf{17.0}   &  40.6   &  49.3   & 52.9  \\ 
Rank+O\textsubscript{\scriptsize{GPT4}}    &  77.4   &  \textbf{17.7}   &  55.6   &  70.0   &  \textbf{82.1}   &  \textbf{88.0}   &  38.6   &  \textbf{17.0}   &  \underline{\textcolor{red}{\textbf{63.4}}}   &  \textbf{71.1}   &  63.4 \\ 
\midrule
\multicolumn{12}{c}{\textbf{\footnotesize{Scoring-based LLM IR Systems}}} \\ 
Score+O\textsubscript{\scriptsize{StripedHyena}}    &  67.5  &  9.5  &  21.6  &  23.1  &  43.0  &  50.8  &  31.8  &  10.8  &  21.1  &  24.8  &  37.0 \\ 
Reason+O\textsubscript{\scriptsize{StripedHyena}}    &  65.9  &  10.9  &  27.7  &  32.0  &  23.1  &  23.1  &  29.2  &  13.8  &  12.7  &  11.9  &  31.7 \\ 
Subtask+O\textsubscript{\scriptsize{StripedHyena}}    &  74.8  &  15.5  &  16.3  &  20.0  &  43.5  &  53.0  &  32.3  &  10.9  &  24.7  &  29.1  &  38.3 \\ 
\\ 
Score+O\textsubscript{\scriptsize{Llama2-7b}}    &  71.6  &  14.1  &  29.3  &  40.1  &  30.5  &  34.7  &  31.1  &  11.2  &  10.9  &  14.0  &  34.7 \\ 
Reason+O\textsubscript{\scriptsize{Llama2-7b}}    &  72.1  &  13.8  &  21.9  &  27.0  &  21.4  &  28.1  &  29.9  &  9.7  &  10.1  &  10.0  &  31.1 \\ 
Subtask+O\textsubscript{\scriptsize{Llama2-7b}}    &  69.2  &  11.8  &  22.3  &  27.9  &  29.2  &  31.1  &  28.7  &  10.0  &  9.4  &  11.9  &  31.8 \\ 
\\ 
Score+O\textsubscript{\scriptsize{Llama2-13b}}    &  77.3  &  14.5  &  15.0  &  24.1  &  29.6  &  34.0  &  30.6  &  10.5  &  16.2  &  18.0  &  33.7 \\ 
Reason+O\textsubscript{\scriptsize{Llama2-13b}}    &  60.1  &  7.5  &  32.6  &  45.1  &  19.4  &  28.0  &  25.4  &  8.7  &  9.1  &  8.9  &  29.3 \\ 
Subtask+O\textsubscript{\scriptsize{Llama2-13b}}    &  71.3  &  14.1  &  18.1  &  20.8  &  52.8  &  61.9  &  27.7  &  10.7  &  19.7  &  20.0  &  37.9 \\ 
\\ 
Score+O\textsubscript{\scriptsize{Llama2-70b}}    &  75.1  &  15.1  &  41.6  &  53.0  &  60.8  &  70.0  &  36.3  &  12.6  &  11.7  &  14.9  &  45.1 \\ 
Reason+O\textsubscript{\scriptsize{Llama2-70b}}    &  70.1  &  14.5  &  45.0  &  60.1  &  43.4  &  48.0  &  32.6  &  11.6  &  12.1  &  15.9  &  40.6 \\ 
Subtask+O\textsubscript{\scriptsize{Llama2-70b}}    &  76.7  &  16.3  &  21.0  &  25.0  &  42.6  &  50.2  &  33.3  &  11.8  &  17.3  &  21.0  &  38.2 \\ 
\\ 
Score+O\textsubscript{\scriptsize{GPT4}}    &  \underline{\textcolor{red}{\textbf{79.9}}}  &  \underline{\textcolor{red}{\textbf{19.3}}}  &  53.2  &  70.8  &  \underline{\textcolor{red}{\textbf{83.3}}}  &  \underline{\textcolor{red}{\textbf{90.9}}}  &  \textbf{43.7}  &  \underline{\textcolor{red}{\textbf{19.6}}}  &  \textbf{56.9}  &  \underline{\textcolor{red}{\textbf{73.2}}}  &  63.4 \\ 
Reason+O\textsubscript{\scriptsize{GPT4}}    &  74.9  &  \textbf{18.1}  &  61.5  &  \textbf{76.1}  &  75.2  &  82.8  &  \underline{\textcolor{red}{\textbf{44.6}}}  &  16.4  &  41.7  &  52.8  &  59.6 \\ 
Subtask+O\textsubscript{\scriptsize{GPT4}}    &  \textbf{78.5}  &  \textbf{18.5}  & \underline{\textcolor{red}{\textbf{70.8}}} &  \underline{\textcolor{red}{\textbf{83.9}}} &  77.9 &  \textbf{85.9} &  \textbf{44.3} &  \textbf{19.5} &  53.3  &  \textbf{66.0} &  \underline{\textcolor{red}{\textbf{65.0}}}\\ 
\bottomrule

\end{tabular}

\caption{nDCG@10 and Recall@5 for the benchmark datasets. \textbf{Bold} indicates no statistically significant difference from the \underline{\textcolor{red}{highest numerical value}}, which is indicated in red. Refer to Appendix Table \ref{tab: model_performance_combined_app} for standard errors. The notation $+$O indicates task objective awareness. The consistently low recall on DORIS-MAE and Clinical-Trial corresponds to their low maximum possible recall.}

\label{tab: model_performance_combined}

\end{table*}

We use BIRCO to evaluate the performance of the different LLM retrieval strategies described in Section \ref{sec: llm-framework}. The strategies are evaluated on GPT4 \cite{openai2023gpt4}, Llama-2 \cite{llama2}, and StripedHyena \cite{StripedHyena}, all of which use the same prompt strategies described in Appendix \ref{sec: appendix_task objective for birco} and Appendix \ref{sec: appendix_prompt}.

Appendix \ref{ssec: appendix_full experiment results} provides details about LLM models, prompt configuration, compute requirements, and additional experimental setups for (fine-tuned) embedding models and fine-tuned decoder models.

Results are shown in Table \ref{tab: model_performance_combined}. Overall, larger LLMs perform better than smaller LLMs, with GPT4 achieving the strongest results across all retrieval strategies. Detailed comparisons of the LLM approaches are provided below.

Table \ref{tab: model_performance_combined} also shows results for the baseline embedding models and language models which have been fine-tuned for information retrieval tasks. Only GPT4-based systems consistently outperform the strongest baselines, RankLLaMA (a 7B model fine-tuned on MS MARCO) and MonoT5 (a 3B model fine-tuned on MS MARCO). Llama 2 70B is the only other LLM that performs better than the E5-L-v2 baseline, which is a 300M parameter embedding model.

Overall, the results show that small embedding models and fine-tuned language models provide strong baselines relative to all LLMs except GPT4. 

\textbf{Ranking vs. Scoring}
We first evaluate the performance of systems that rank documents by comparing them, and systems that directly score the relevance of individual documents. We find no general effect of comparative ranking vs. direct scoring, except for Llama 70B, which had stronger performance with direct scoring.

\textbf{Chain-of-Thought Reasoning}
We next evaluate whether chain-of-thought reasoning improves retrieval performance for LLMs. While chain-of-thought improves performance for certain task/model combinations, the effect is inconsistent, and in aggregate chain-of-thought slightly decreases retrieval performance. 

While chain-of-thought frequently improves performance on NLP and reasoning tasks, other work has shown that improvements are not uniformly found across all tasks \citep{Big_Bench_hard}, consistent with the current results. 

\textbf{Task Decomposition}
With task decomposition, the model first decomposes the query into subtasks, and then evaluates each subtask separately. Task decomposition does not lead to consistent improvement in performance. The strongest average performance is achieved by GPT4 with task decomposition, but this improvement is largely driven by a single task, ArguAna.

\textbf{Task Objective Awareness}
Providing models with detailed descriptions of task objectives does not improve performance on the smaller LLMs. For GPT4 with comparative ranking, it substantially increases performance on two datasets, ArguAna and RELIC. For GPT4 with direct scoring, we found poor performance on the development sets when task objectives were not provided (results not shown in Table \ref{tab: model_performance_combined}).
Qualitatively, this was due to the model not knowing how to determine relevance between queries and passages. 

ArguAna and RELIC have non-standard task objectives, likely accounting for the importance of detailed prompting for these tasks.

\subsection{Error Analysis}
Two of the datasets contain hard negatives, non-relevant documents which are constructed to be similar to the ground-truth answers, as described in Section \ref{sec:hard-negatives}. Table \ref{tab: compact_arr} shows the average reciprocal rank for hard negatives compared to the other passages in the datasets, as ranked by GPT4-based systems. The reciprocal rank of the hard negatives is much higher than the other passages, indicating that GPT4 found these passages more relevant, and that these passages increase the difficulty of retrieval. See more details of error analysis in Appendix \ref{sec: appendix_error analysis for models}.
\begin{table}[ht]
\centering
\setlength{\tabcolsep}{1.5pt}
\begin{tabular}{lcc|cc}
\toprule
\textbf{Method} & \multicolumn{2}{c|}{\textbf{Clinical-Trial}} & \multicolumn{2}{c}{\textbf{ArguAna}} \\
 & Hard & Other & Hard & Other \\
\midrule
Reason+O\textsubscript{\scriptsize{GPT4}} & .097 & .042 & .154 & .059 \\
Score+O\textsubscript{\scriptsize{GPT4}} & .097 & .040 & .170 & .056 \\
Subtask+O\textsubscript{\scriptsize{GPT4}} & .090 & .042 & .148 & .057 \\
RankGPT+O\textsubscript{\scriptsize{GPT4}} & .096 & .049 & .166 & .056 \\
\midrule
\textbf{Avg.} & .095 & .043 & .160 & .057 \\
\bottomrule
\end{tabular}

\caption{Average reciprocal rank for hard negatives. }
\label{tab: compact_arr}
\end{table}

\section{Conclusion}

We have introduced BIRCO, a benchmark for IR tasks with complex objectives. BIRCO includes 5 diverse retrieval tasks, and is significantly more challenging than previous IR benchmarks.

We found that embedding methods and small language models have weak performance on the BIRCO tasks. Methods that use LLMs for ranking have stronger performance, though none achieve strong results across all tasks.

We evaluated several hypotheses regarding LLM performance. First, providing clear instructions to the LLMs regarding task objectives improved performance on certain tasks. Second, we did not find evidence that ranking by comparing passages improved performance relative to directly scoring passages. Third, in contrast to results in many other NLP domains \cite{cot1,cot4}, we did not find evidence that chain-of-thought reasoning improves retrieval performance. Finally, decomposing queries into subtasks improved performance on only a single dataset.

The results underscore the need to develop IR methods that go beyond similarity-based retrieval. Strong performance on BIRCO requires models that can understand multi-faceted user intents. While GPT4-based methods had the strongest performance, even they did not achieve adequate performance across tasks. Furthermore, it is currently prohibitively expensive to perform inference with LLMs for all but the smallest IR tasks. The challenge of complex user objectives will require improvements in model abilities and efficiency.

\newpage

\section{Limitations}

There are many types of user retrieval needs. BIRCO only provides a benchmark for a limited sample of these needs. We hope that future datasets and benchmarks can provide more comprehensive coverage of these other cases. 

Large language models are computationally expensive and can only be effectively employed in the passage re-ranking stage of a multi-stage IR process. In addition, depending on the use case, LLMs may be prohibitively expensive even when the candidate pool size is limited.

While we optimized prompts to the best of our abilities on the development sets, it is possible that better prompts would improve LLM performance on the tasks.

\section{Ethical Considerations}

No ethical concerns for this work.

\newpage
\bibliography{main}

\newpage
\appendix
\section{Experiment Details}
\label{sec: appendix_experiment}

\begin{table*}
\scriptsize
\setlength{\tabcolsep}{2pt}
\centering
\begin{tabular}{l cc | cc | cc | cc | cc | C{1cm}}
\toprule
\textbf{\footnotesize{Model}} & \multicolumn{2}{c}{\textbf{\footnotesize{DORIS-MAE}}} & \multicolumn{2}{c}{\textbf{\footnotesize{ArguAna}}} & \multicolumn{2}{c}{\textbf{\footnotesize{WhatsThatBook}}} & \multicolumn{2}{c}{\textbf{\footnotesize{Clinical-Trial}}} & \multicolumn{2}{c}{\textbf{\footnotesize{RELIC}}}  \\
& \scriptsize{nDCG@10} & \scriptsize{Recall@5} & \scriptsize{nDCG@10} & \scriptsize{Recall@5} & \scriptsize{nDCG@10} & \scriptsize{Recall@5} & \scriptsize{nDCG@10} & \scriptsize{Recall@5} & \scriptsize{nDCG@10} & \scriptsize{Recall@5} & \scriptsize{Avg. nDCG@10} \\
\midrule
Maximum Possible & 100.0 & 45.6 & 100.0 & 100.0 & 100.0 & 100.0 & 100.0 & 43.0 & 100.0 & 100.0 & 100.0 \\
Random & 49.9 & 4.5 & 9.1  & 9.9  & 8.9 & 9.8 & 22.4  & 8.4  & 8.9  & 10.0 & 19.8 \\
\midrule
\multicolumn{12}{c}{\textbf{\footnotesize{Embedding Models}}} \\ 
RoBERTa-L    &  66.8 $\pm$ 1.3  &  12.0 $\pm$ 1.6  &  31.8 $\pm$ 3.5  &  41.2 $\pm$ 4.8  &  14.6 $\pm$ 2.7  &  16.0 $\pm$ 3.7  &  26.6 $\pm$ 2.3  &  9.4 $\pm$ 1.4  &  7.0 $\pm$ 1.8  &  7.8 $\pm$ 2.6  &  29.4 \\ 
SPLADE++    &  66.8 $\pm$ 1.3  &  8.2 $\pm$ 1.2  &  34.2 $\pm$ 3.3  &  47.6 $\pm$ 4.9  &  9.5 $\pm$ 2.2  &  8.9 $\pm$ 2.9  &  28.0 $\pm$ 2.2  &  10.8 $\pm$ 1.5  &  10.4 $\pm$ 2.1  &  12.9 $\pm$ 3.4  &  29.8 \\ 
SPLADE-v2    &  67.9 $\pm$ 1.4  &  10.6 $\pm$ 2.1  &  38.5 $\pm$ 3.6  &  41.9 $\pm$ 4.8  &  16.2 $\pm$ 3.0  &  20.2 $\pm$ 4.1  &  23.0 $\pm$ 2.3  &  7.5 $\pm$ 1.2  &  10.4 $\pm$ 2.2  &  11.0 $\pm$ 3.1  &  31.2 \\ 
E5-L-v2    &  72.0 $\pm$ 1.2  &  14.6 $\pm$ 1.9  &  44.9 $\pm$ 3.2  &  62.0 $\pm$ 4.8  &  36.6 $\pm$ 4.0  &  39.9 $\pm$ 4.8  &  30.4 $\pm$ 2.6  &  11.3 $\pm$ 1.8  &  10.9 $\pm$ 2.5  &  14.9 $\pm$ 3.6  &  39.0 \\ 
SIMCSE    &  70.8 $\pm$ 1.3  &  14.4 $\pm$ 1.8  &  40.3 $\pm$ 3.5  &  51.9 $\pm$ 5.1  &  31.7 $\pm$ 4.0  &  37.0 $\pm$ 5.0  &  27.1 $\pm$ 2.8  &  10.7 $\pm$ 1.6  &  12.7 $\pm$ 2.4  &  15.9 $\pm$ 3.6  &  36.5 \\ 
Promptagator  & 76.0 $\pm$ 1.4 & 15.6 $\pm$ 1.8 & 60.9 $\pm$3.2  & 73.7 $\pm$ 4.4 & 58.4 $\pm$ 4.2 & 62.9 $\pm$ 4.7 & NA & NA & NA & NA & NA\\
\midrule
\multicolumn{12}{c}{\textbf{\footnotesize{Finetuned Encoder-Decoder Models}}} \\ 
TART    &  64.9 $\pm$ 1.5  &  10.2 $\pm$ 1.7  &  49.3 $\pm$ 3.4  &  66.7 $\pm$ 4.8  &  42.7 $\pm$ 3.7  &  46.1 $\pm$ 4.9  &  32.9 $\pm$ 2.7  &  11.1 $\pm$ 1.4  &  9.1 $\pm$ 1.7  &  7.0 $\pm$ 2.5  &  39.8 \\ 
TART+O    &  49.6 $\pm$ 1.8  &  4.9 $\pm$ 1.0  &  23.6 $\pm$ 2.8  &  29.9 $\pm$ 4.5  &  17.9 $\pm$ 2.4  &  20.7 $\pm$ 4.1  &  24.1 $\pm$ 2.3  &  7.5 $\pm$ 1.1  &  13.2 $\pm$ 2.2  &  17.1 $\pm$ 3.7  &  25.7 \\ 
MonoT5    &  66.9 $\pm$ 1.5  &  12.4 $\pm$ 1.8  &  46.8 $\pm$ 3.4  &  52.7 $\pm$ 5.1  &  50.8 $\pm$ 4.2  &  59.0 $\pm$ 4.9  &  34.3 $\pm$ 2.5  &  15.6 $\pm$ 2.5  &  11.9 $\pm$ 2.5  &  10.0 $\pm$ 3.0  &  42.1 \\ 
\midrule
\multicolumn{12}{c}{\textbf{\footnotesize{Finetuned Decoder-Only Model}}} \\ 
RankLLaMA    &  75.1 $\pm$ 1.3  &  18.4 $\pm$ 2.0  &  54.1 $\pm$ 3.0  &  72.6 $\pm$ 4.5  &  63.7 $\pm$ 3.9  &  69.8 $\pm$ 4.6  &  30.7 $\pm$ 2.2  &  10.9 $\pm$ 1.4  &  14.9 $\pm$ 2.4  &  18.0 $\pm$ 3.7  &  47.7 \\ 
\midrule
\multicolumn{12}{c}{\textbf{\footnotesize{Comparison-based LLM IR Systems}}} \\ 
Rank\textsubscript{\scriptsize{SH}}    &  71.7 $\pm$ 1.3  &  14.6 $\pm$ 1.8  &  39.8 $\pm$ 3.0  &  58.0 $\pm$ 4.9  &  33.6 $\pm$ 3.8  &  39.8 $\pm$ 4.8  &  29.5 $\pm$ 2.2  &  10.3 $\pm$ 1.5  &  10.9 $\pm$ 2.5  &  13.0 $\pm$ 3.3  &  37.1 \\ 
Rank+O\textsubscript{\scriptsize{SH}}    &  70.6 $\pm$ 1.3  &  15.5 $\pm$ 1.9  &  38.9 $\pm$ 3.0  &  54.8 $\pm$ 5.0  &  32.8 $\pm$ 3.8  &  39.9 $\pm$ 4.8  &  29.8 $\pm$ 2.4  &  10.8 $\pm$ 1.5  &  9.2 $\pm$ 2.0  &  11.9 $\pm$ 3.2  &  36.3 \\ 
\\ 
Rank\textsubscript{\scriptsize{LM2-7b}}    &  71.0 $\pm$ 1.3  &  13.2 $\pm$ 1.6  &  40.5 $\pm$ 3.1  &  58.9 $\pm$ 4.8  &  25.7 $\pm$ 2.8  &  35.9 $\pm$ 4.7  &  31.2 $\pm$ 2.6  &  11.5 $\pm$ 1.6  &  11.1 $\pm$ 2.5  &  11.9 $\pm$ 3.2  &  35.9 \\ 
Rank+O\textsubscript{\scriptsize{LM2-7b}}    &  70.8 $\pm$ 1.3  &  14.1 $\pm$ 1.7  &  39.4 $\pm$ 2.9  &  50.9 $\pm$ 4.9  &  26.5 $\pm$ 2.9  &  37.8 $\pm$ 4.8  &  28.5 $\pm$ 2.5  &  9.6 $\pm$ 1.5  &  10.8 $\pm$ 2.4  &  14.0 $\pm$ 3.5  &  35.2 \\ 
\\ 
Rank\textsubscript{\scriptsize{LM2-13b}}    &  71.2 $\pm$ 1.4  &  13.0 $\pm$ 1.6  &  41.1 $\pm$ 2.9  &  57.0 $\pm$ 4.8  &  27.1 $\pm$ 3.0  &  37.8 $\pm$ 4.9  &  32.9 $\pm$ 3.0  &  13.8 $\pm$ 2.3  &  9.9 $\pm$ 2.0  &  14.9 $\pm$ 3.6  &  36.4 \\ 
Rank+O\textsubscript{\scriptsize{LM2-13b}}    &  70.5 $\pm$ 1.3  &  14.2 $\pm$ 1.7  &  37.8 $\pm$ 2.9  &  47.8 $\pm$ 5.0  &  28.0 $\pm$ 3.1  &  38.9 $\pm$ 4.9  &  30.0 $\pm$ 2.5  &  10.9 $\pm$ 1.5  &  9.2 $\pm$ 2.1  &  12.9 $\pm$ 3.3  &  35.1 \\ 
\\ 
Rank\textsubscript{\scriptsize{LM2-70b}}    &  70.7 $\pm$ 1.3  &  15.1 $\pm$ 1.9  &  41.3 $\pm$ 3.1  &  55.8 $\pm$ 4.9  &  30.2 $\pm$ 3.3  &  37.8 $\pm$ 4.8  &  29.0 $\pm$ 2.5  &  9.8 $\pm$ 1.5  &  10.6 $\pm$ 2.4  &  12.9 $\pm$ 3.4  &  36.4 \\ 
Rank+O\textsubscript{\scriptsize{LM2-70b}}    &  70.4 $\pm$ 1.3  &  14.2 $\pm$ 1.7  &  40.2 $\pm$ 3.1  &  55.9 $\pm$ 5.0  &  25.6 $\pm$ 2.9  &  38.9 $\pm$ 4.8  &  30.0 $\pm$ 2.5  &  11.4 $\pm$ 1.8  &  9.1 $\pm$ 2.0  &  10.9 $\pm$ 3.1  &  35.1 \\ 
\\ 
Rank\textsubscript{\scriptsize{GPT4}}    &  76.2 $\pm$ 1.3  &  17.5 $\pm$ 2.3  &  27.3 $\pm$ 2.2  &  17.0 $\pm$ 3.7  &  80.1 $\pm$ 3.4  &  86.9 $\pm$ 3.3  &  40.3 $\pm$ 2.7  &  17.0 $\pm$ 2.4  &  40.6 $\pm$ 3.9  &  49.3 $\pm$ 4.8  & 52.9  \\ 
Rank+O\textsubscript{\scriptsize{GPT4}}    &  77.4 $\pm$ 1.3  &  17.7 $\pm$ 2.1  &  55.6 $\pm$ 3.4  &  70.0 $\pm$ 4.7  &  82.1 $\pm$ 3.3  &  88.0 $\pm$ 3.2  &  38.6 $\pm$ 2.9  &  17.0 $\pm$ 2.5  &  63.4 $\pm$ 4.0  &  71.1 $\pm$ 4.6  &  63.4 \\ 
\midrule
\multicolumn{12}{c}{\textbf{\footnotesize{Scoring-based LLM IR Systems}}} \\ 
Score+O\textsubscript{\scriptsize{SH}}    &  67.5 $\pm$ 1.4  &  9.5 $\pm$ 1.4  &  21.6 $\pm$ 2.6  &  23.1 $\pm$ 4.2  &  43.0 $\pm$ 4.0  &  50.8 $\pm$ 4.9  &  31.8 $\pm$ 2.7  &  10.8 $\pm$ 1.5  &  21.1 $\pm$ 3.0  &  24.8 $\pm$ 4.4  &  37.0 \\ 
Reason+O\textsubscript{\scriptsize{SH}}    &  65.9 $\pm$ 1.1  &  10.9 $\pm$ 1.4  &  27.7 $\pm$ 3.5  &  32.0 $\pm$ 4.6  &  23.1 $\pm$ 3.7  &  23.1 $\pm$ 4.4  &  29.2 $\pm$ 2.8  &  13.8 $\pm$ 2.5  &  12.7 $\pm$ 2.4  &  11.9 $\pm$ 3.2  &  31.7 \\ 
Subtask+O\textsubscript{\scriptsize{SH}}    &  74.8 $\pm$ 1.3  &  15.5 $\pm$ 2.3  &  16.3 $\pm$ 2.2  &  20.0 $\pm$ 4.0  &  43.5 $\pm$ 3.7  &  53.0 $\pm$ 5.0  &  32.3 $\pm$ 2.9  &  10.9 $\pm$ 1.7  &  24.7 $\pm$ 3.2  &  29.1 $\pm$ 4.4  &  38.3 \\ 
\\ 
Score+O\textsubscript{\scriptsize{LM2-7b}}    &  71.6 $\pm$ 1.4  &  14.1 $\pm$ 1.9  &  29.3 $\pm$ 3.2  &  40.1 $\pm$ 4.8  &  30.5 $\pm$ 3.8  &  34.7 $\pm$ 4.8  &  31.1 $\pm$ 2.7  &  11.2 $\pm$ 1.6  &  10.9 $\pm$ 2.4  &  14.0 $\pm$ 3.5  &  34.7 \\ 
Reason+O\textsubscript{\scriptsize{LM2-7b}}    &  72.1 $\pm$ 1.2  &  13.8 $\pm$ 1.7  &  21.9 $\pm$ 3.4  &  27.0 $\pm$ 4.5  &  21.4 $\pm$ 3.1  &  28.1 $\pm$ 4.4  &  29.9 $\pm$ 2.7  &  9.7 $\pm$ 1.5  &  10.1 $\pm$ 2.2  &  10.0 $\pm$ 3.1  &  31.1 \\ 
Subtask+O\textsubscript{\scriptsize{LM2-7b}}    &  69.2 $\pm$ 1.4  &  11.8 $\pm$ 1.7  &  22.3 $\pm$ 2.5  &  27.9 $\pm$ 4.5  &  29.2 $\pm$ 3.5  &  31.1 $\pm$ 4.5  &  28.7 $\pm$ 2.4  &  10.0 $\pm$ 1.2  &  9.4 $\pm$ 2.2  &  11.9 $\pm$ 3.2  &  31.8 \\ 
\\ 
Score+O\textsubscript{\scriptsize{LM2-13b}}    &  77.3 $\pm$ 0.9  &  14.5 $\pm$ 1.6  &  15.0 $\pm$ 2.2  &  24.1 $\pm$ 4.2  &  29.6 $\pm$ 3.4  &  34.0 $\pm$ 4.6  &  30.6 $\pm$ 2.6  &  10.5 $\pm$ 1.3  &  16.2 $\pm$ 2.6  &  18.0 $\pm$ 3.9  &  33.7 \\ 
Reason+O\textsubscript{\scriptsize{LM2-13b}}    &  60.1 $\pm$ 1.4  &  7.5 $\pm$ 1.4  &  32.6 $\pm$ 3.1  &  45.1 $\pm$ 5.0  &  19.4 $\pm$ 2.4  &  28.0 $\pm$ 4.5  &  25.4 $\pm$ 2.1  &  8.7 $\pm$ 1.2  &  9.1 $\pm$ 1.9  &  8.9 $\pm$ 2.7  &  29.3 \\ 
Subtask+O\textsubscript{\scriptsize{LM2-13b}}    &  71.3 $\pm$ 1.3  &  14.1 $\pm$ 1.7  &  18.1 $\pm$ 2.7  &  20.8 $\pm$ 4.1  &  52.8 $\pm$ 3.9  &  61.9 $\pm$ 5.0  &  27.7 $\pm$ 2.5  &  10.7 $\pm$ 1.4  &  19.7 $\pm$ 2.8  &  20.0 $\pm$ 4.0  &  37.9 \\ 
\\ 
Score+O\textsubscript{\scriptsize{LM2-70b}}    &  75.1 $\pm$ 1.2  &  15.1 $\pm$ 2.0  &  41.6 $\pm$ 3.4  &  53.0 $\pm$ 4.9  &  60.8 $\pm$ 3.7  &  70.0 $\pm$ 4.7  &  36.3 $\pm$ 2.8  &  12.6 $\pm$ 1.7  &  11.7 $\pm$ 2.5  &  14.9 $\pm$ 3.6  &  45.1 \\ 
Reason+O\textsubscript{\scriptsize{LM2-70b}}    &  70.1 $\pm$ 1.4  &  14.5 $\pm$ 1.9  &  45.0 $\pm$ 3.3  &  60.1 $\pm$ 4.8  &  43.4 $\pm$ 4.1  &  48.0 $\pm$ 5.0  &  32.6 $\pm$ 2.5  &  11.6 $\pm$ 1.5  &  12.1 $\pm$ 2.5  &  15.9 $\pm$ 3.7  &  40.6 \\ 
Subtask+O\textsubscript{\scriptsize{LM2-70b}}    &  76.7 $\pm$ 1.2  &  16.3 $\pm$ 1.9  &  21.0 $\pm$ 2.9  &  25.0 $\pm$ 4.4  &  42.6 $\pm$ 3.7  &  50.2 $\pm$ 5.0  &  33.3 $\pm$ 3.0  &  11.8 $\pm$ 1.8  &  17.3 $\pm$ 2.8  &  21.0 $\pm$ 4.0  &  38.2 \\ 
\\ 
Score+O\textsubscript{\scriptsize{GPT4}}    &  79.9 $\pm$ 1.2  &  19.3 $\pm$ 2.0  &  53.2 $\pm$ 2.8  &  70.8 $\pm$ 4.6  &  83.3 $\pm$ 3.1  &  90.9 $\pm$ 2.8  &  43.7 $\pm$ 2.6  &  19.6 $\pm$ 2.5  &  56.9 $\pm$ 3.2  &  73.2 $\pm$ 4.3  &  63.4 \\ 
Reason+O\textsubscript{\scriptsize{GPT4}}    &  74.9 $\pm$ 1.3  &  18.3 $\pm$ 2.3  &  61.5 $\pm$ 3.0  &  76.1 $\pm$ 4.4  &  75.2 $\pm$ 3.6  &  82.8 $\pm$ 3.7  &  44.6 $\pm$ 3.1  &  16.4 $\pm$ 1.9  &  41.7 $\pm$ 3.3  &  52.8 $\pm$ 4.9  &  59.6 \\ 
Subtask+O\textsubscript{\scriptsize{GPT4}}    &  78.5 $\pm$ 1.2  &  18.5 $\pm$ 1.9  &  70.8 $\pm$ 3.1  &  83.9 $\pm$ 3.6  &  77.9 $\pm$ 3.5  &  85.9 $\pm$ 3.4  &  44.3 $\pm$ 2.9  &  19.5 $\pm$ 2.5  &  53.3 $\pm$ 3.7  &  66.0 $\pm$ 4.6  &  65.0\\ 
\bottomrule
\end{tabular}
\caption{nDCG@10 and Recall@5 for the benchmark datasets. 
Error bars are standard errors. The notation $+$O indicates task objective awareness. LM2 stands for Llama2 \cite{llama2}. SH stands for StripedHyena \cite{StripedHyena}}
\label{tab: model_performance_combined_app}
\end{table*}

We use the GPT4-0613 checkpoint for all GPT4-based systems and access it via OpenAI's API. We use Together AI's API to access the other LLMs' checkpoints \cite{together_ai}.

Each GPT4-based system takes less than 12 hours to evaluate on BIRCO, and costs approximately \$500-\$1000. Other LLM-based systems take less than 3 hours to evaluate on BIRCO and cost less than \$50. 

All embedding models have 350M parameters or less. Promptagator models are E5-v2-L checkpoints finetuned on promptagator-styled \cite{promptagator} synthetic data, introduced by \cite{ir2}. No data for Clinical-Trial and RELIC is available for Promptagator models. Other embedding models are not fine-tuned for IR tasks.

Finetuning for TART, MonoT5, and RankLLama was performed on MS MARCO. Experiments on these models can be run on one node of an 8 NVIDIA H100 GPU (80G) within one hour. 

\subsection{Baselines}
\label{ssec: appendix_baselines}
For pretrained embedding models, we choose several recent state-of-the-art models as well as several that have been extensively benchmarked on other tasks: E5-Large-v2 \cite{e5}, SimCSE-Large \cite{simcse}, SPECTER-v2 \cite{specterv2}, ROBERTA-Large \cite{roberta}, SPLADE-v2 \cite{spladev2}, and SPLADE++ \cite{splade++}. There are several encoder-decoder models specifically trained for information retrieval: monoT5-3B \cite{monot5}, which is supervised on MS MARCO, and TART-1.5B \cite{asai-etal-2023-task}, which is instruction-tuned and can incorporate task descriptions. For DORIS-MAE, ArguAna, and WhatsThatBook, we also report results for E5-Large-v2 checkpoints fine-tuned using synthetic data generated by Promptagator \cite{promptagator} from \cite{ir2}. Additionally, we evaluate a decoder-only model RankLLaMA-7B \cite{rankllama} trained on MS MARCO.
We also evaluate two other generative models: Llama2 chat \cite{llama2}, which is one of the best open-sourced LLMs, and StripedHyena-Nous-7B chat \cite{StripedHyena}, which is a competitive alternative LLM with a hybrid architecture between multi-headed attention and gated convolution. 

\subsection{Full Experiment Results}
\label{ssec: appendix_full experiment results}
See Table \ref{tab: model_performance_combined_app} for full experiment results with more embedding models and standard errors for each model's performance. 
\subsubsection{Analysis for Non-LLM models}
As shown, all embedding models are significantly worse than GPT4-based models. On average, fine-tuned decoder models TART and MonoT5 are comparable with the best pretrained embedding models. The larger RankLLaMA decoder model outperforms most embedding models and other decoder models. However, based on Promptagator's performance on DORIS-MAE, ArguAna, and WhatsThatBook, this synthetic data trained embedding model appears comparable to RankLLaMA. There is a significant decrease in TART+O compared with TART, suggesting TART has difficulty following complex instructions.

\section{Task Objectives for BIRCO}
\label{sec: appendix_task objective for birco}

\begin{figure*}[ht]
    \centering
    \includegraphics[width=\linewidth]{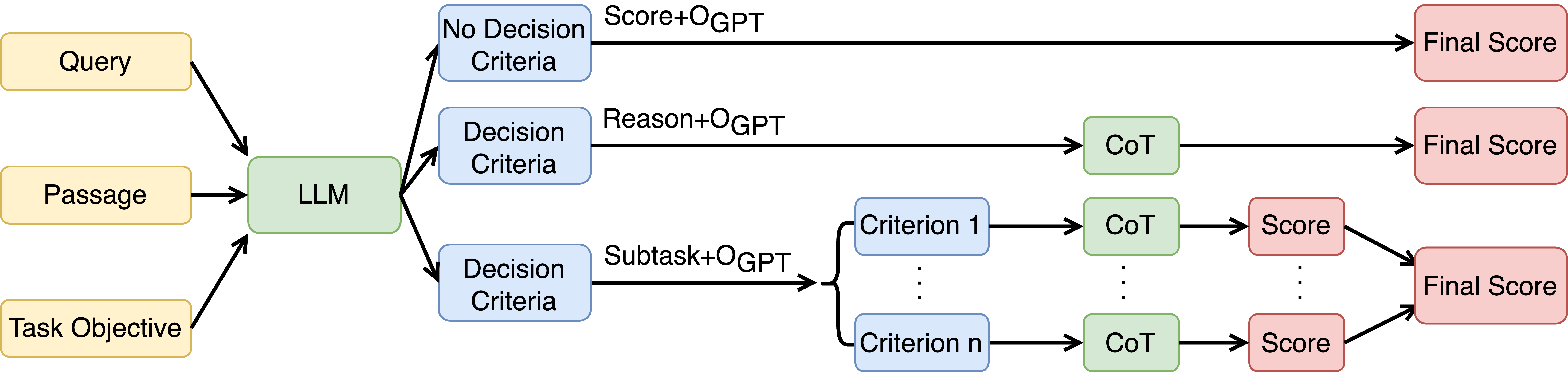}
    \caption{A framework for integrating LLMs into retrieval tasks}
    \label{fig: SER_DeAGO_Framework}
\end{figure*}

\begin{figure*}[ht]
    \centering
    \includegraphics[width=\linewidth]{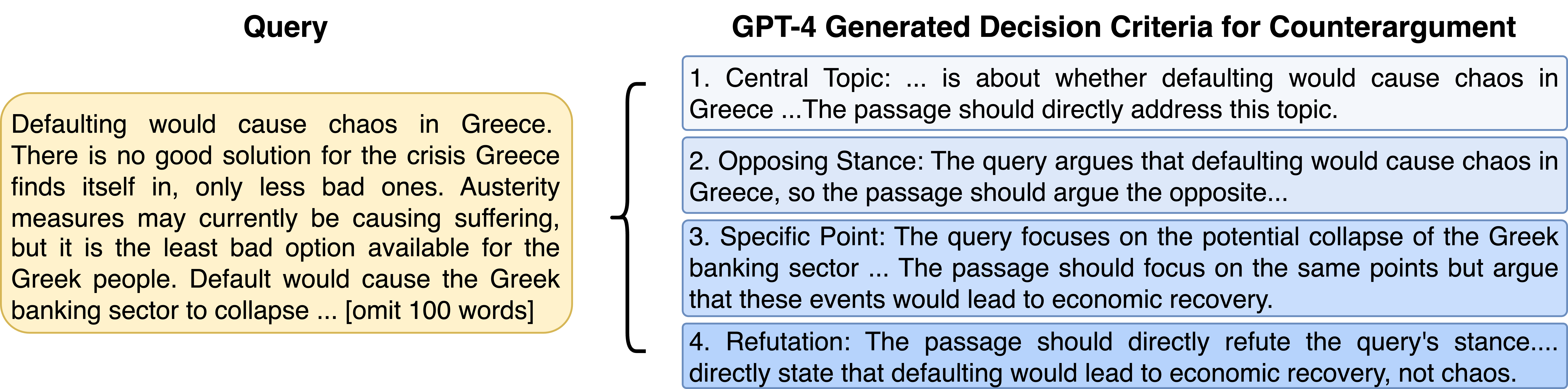}
    \caption{An example query and LLM-generated decision criteria}
    \label{fig: example_decision_criteria}
\end{figure*}

\begin{table*}[htbp]
\centering
\setlength{\tabcolsep}{1.7pt}
\footnotesize
\begin{tabular}{llcccccccc}
\toprule
Dataset & Dataset Name & R/Q & Corpus Size & Relevancy & Recall@5 & Recall@20 & nDCG@10 & MRR@10 & MAP \\
\midrule
\multirow{5}{*}{BIRCO} & DORIS-MAE & 18.2 & 110.6 & Multi-level & 14.7 & 42.2 & 72.0 & 14.3 & 40.5 \\
 & ArguAna & 1.0 & 50.0 & Binary & 62.0 & 96.0 & 43.6 & 33.2 & 34.7 \\
 & WhatsThatBook & 1.0 & 50.4 & Binary & 40.0 & 64.0 & 36.8 & 32.0 & 33.8 \\
 & Clinical-Trial & 20.9 & 68.4 & 3-level & 10.6 & 37.8 & 29.4 & 34.4 & 37.4 \\
 & RELIC & 1.0 & 50.6 & Binary & 15.0 & 43.0 & 11.2 & 8.9 & 12.2 \\
 & \textbf{Average} & 8.4 & 66.0 & - & 28.5 & 56.6 & \textbf{38.6} & 24.6 & 31.7 \\
 \midrule

 \multirow{14}{*}{BEIR} & MS MARCO & 1.1 & 8842K & Binary & - & - & 44.1 & - & - \\
 & TREC-COVID & 493.5 & 171K & 3-level & - & - & 78.3 & - & - \\
 & NFCorpus & 38.2 & 4K & 3-level & - & - & 36.1 & - & - \\
 & NQ & 1.2 & 2681K & Binary & - & - & 62.9 & - & - \\
 & HotpotQA & 2.0 & 5233K & Binary & - & - & 63.3 & - & - \\
 & FiQA & 2.6 & 58K & Binary & - & - & 38.6 & - & - \\
 & Touche & 19.0 & 383K & 3-level & - & - & 27.2 & - & - \\
 & CQADupStack & 1.4 & 457K & Binary & - & - & 39.4 & - & - \\
 & Quora & 1.6 & 523K & Binary & - & - & 88.2 & - & - \\
 & DBPedia & 38.2 & 4636K & 3-level & - & - & 42.4 & - & - \\
 & SCIDOCS & 4.9 & 26K & Binary & - & - & 20.1 & - & - \\
 & FEVER & 1.2 & 5417K & Binary & - & - & 65.0 & - & - \\
 & Climate-FEVER & 3.0 & 5417K & Binary & - & - & 22.4 & - & - \\
 & SciFact & 1.1 & 5K & Binary & - & - & 72.6 & - & - \\
 & \textbf{Average} & 43.5 & 2418K & - & - & - & \textbf{50.0} & - & - \\
\bottomrule
\end{tabular}
\caption{Comparison between E5-v2's performances on BIRCO and BEIR. R/D indicates the average number of relevant documents per query. E5-v2's original paper only reports nDCG@10 for BEIR.}
\label{tab: appendix_comparison_between_birco_and_beir}
\end{table*}

We report our prompts that describe task objectives. These prompts were engineered on a small-scale development set for each dataset. The dev set for DORIS-MAE has 40 queries. The dev set for Clinical-Trial has the rest 9 queries. All the other datasets have 50 queries in their dev sets. See extraction procedure for triplets from these queries and their candidate pools in Appendix \ref{sec: appendix_dev_set}.

\subsection{Task Objective for DORIS-MAE}
\label{ssec: appendix_task objective for doris-mae}

"The query consists of users’ needs, leading to several research questions that span a paragraph. Each candidate passage is an abstract from a scientific paper. The objective of this information retrieval task is to identify the abstract that most effectively meets the user's needs in the query."

\subsection{Task Objective for ArguAna}
\label{ssec: appendix task objective for arguana}
"This information retrieval (IR) task has a debate format where a topic is given, and two directly opposing sides of arguments about this topic are formed.  A query is an argument that takes one side of this topic, focuses on a particular point about this topic, and takes a stance (i.e., opinion, position, view, perspective) about this particular point. A passage is an argument that takes the opposing side of the same topic, focuses on the same particular point about the same topic, and takes a directly opposing stance that directly (i.e., no implying or inferring) refutes and attacks the query’s stance regarding this particular point. Both query and passage might have citations in them but these citations should not be considered in the scope of this task. The overall goal of this specific information retrieval IR task is to identify the central topic of the debate, to articulate the query’s stance, and to find the passage that takes the opposing stance."

\subsection{Task Objective for WhatsThatBook}
\label{ssec: appendix_task objective for wtb}

"The query has this format: a user is trying to remember the name of a specific book. The user only remembers some details about the book, such as places, events, and some characters’ names. Some of the details are described using informal language. The passage is a book description or summary of a specific book. The passage typically describes the overall storyline of the book and contains some details about the book. The objective of this information retrieval IR task is for you to find the passage that has details or components that holistically best match, explicitly or implicitly, the details or components raised in the query. In other words, you need to find the book description (i.e., the passage) that is most likely the book the user is looking for in the query."

\subsection{Task Objective for RELIC}
\label{ssec: appendix_task objective for relic}

"The query is a piece of literary analysis written by a scholar. In the query (i.e., the excerpt from a literary analysis), one or more quotations from a classic English novel is used as evidence to support the claims made by the literary analysis. Quotations are identified by quotation marks. Now, one quotation is being intentionally masked from the literary analysis (i.e., the query), and replaced by the symbol [masked sentence(s)]. An important claim is made in the preceding context and another important point is made in the subsequent context surrounding the [masked sentence(s)]. The objective of this information retrieval task is to find the most suitable passage that can be used to **directly support** at least one claim made in the query (i.e., the claim that is made in the preceding or the claim subsequent context surrounding the [masked sentence(s)]) and is very natural to be plugged into the [masked sentence(s)] part of the query. Obviously the most suitable passage should **NOT REPEAT** or be contained in any part of the query. It does not make sense to repeat the same or very similar things twice in literary analysis."

\subsection{Task Objective for Clinical-Trial}
\label{ssec: appendix_task objective for clinical-trial}
"The motivation of the Information Retrieval task is that clinical trials are experiments conducted in the development of new medical treatments, drugs or devices, and recruiting candidates for a trial is often a time-consuming and resource-intensive effort. A query is a patient case report (either in the form of electronic patient records or ad-hoc queries). A passage is a clinical trial. This Information Retrieval task is to improve patient recruitment for clinical trials. The overall goal of this specific information retrieval IR task is to match eligible patients (the query) to clinical trials (the passage) for recruitment."

\section{The Effects of Candidate Pool Construction}
\label{sec: appendix_the effects of candidate pool construction}
In order to make LLM-based systems' evaluation on BIRCO cost-effective, we use the mixture of ada-002 embedding model and BM25 algorithm to construct a candidate pool of size 50 for each query in ArguAna, WhatsThatBook and RELIC. We also randomly sample 100 queries from the above three datasets' decontaminated set of queries, subject to a length constraint (i.e. queries that have fewer than 50 tokens are filtered). For DORIS-MAE, we use the original paper's test set. For Clinical-Trial we use 50 decontaminated queries and their original candidate pools as test set. The remaining 9 queries are used as a dev set.

Please refer to Table \ref{tab: test_set_whole_set_comparison} for the statistics about the test set and the whole dataset. The reason that the ArguAna test set from BIRCO exhibits lower E5 performance than the entire ArguAna dataset is potentially due to shorter queries being filtered out. Longer queries are possibly more challenging queries. As a reference, the average query length for BIRCO's ArguAna is 213 while the average query length for ArguAna's whole set is 193. 

Clinical-Trial test set is competitive with its complete set, which has 9 more queries that include 4 contaminated queries. Potentially contaminated queries (i.e. queries that GPT4 can answer without prior knowledge) are easier.

Conversely, on RELIC and WhatsThatBook, the test sets are easier, as indicated by higher E5 performance. This is due to a significantly smaller candidate pool size, which reduces the difficulty of the task.

\begin{table*}[ht]
\centering
\footnotesize
\setlength{\tabcolsep}{3pt}
\begin{tabular}{@{}lccccc@{}}
\toprule
Dataset & \multicolumn{2}{c}{Test Set} & \phantom{abc} & \multicolumn{2}{c}{Whole Dataset} \\
\cmidrule(lr){2-3} \cmidrule(lr){5-6}
& E5-v2 nDCG@10 & E5-v2 R@5 && E5-v2 nDCG@10 & E5-v2 R@5 \\
\midrule
ArguAna & 44.9  {$\pm$ 3.2} & 62.0  {$\pm$ 4.8} &&  48.0  {$\pm$ 1.0} & 59.7  {$\pm$ 1.4} \\
WhatsThatBook & 36.6  {$\pm$ 4.0} & 39.9  {$\pm$ 4.8} &&   21.0 {$\pm$ 2.8} & 24.7  {$\pm$ 3.5} \\
Clinical-Trial & 30.4 {$\pm$ 2.6}  & 11.3 {$\pm$ 1.8} && 32.4 {$\pm$ 2.6} & 13.1  {$\pm$ 1.9}\\
RELIC & 10.9 {$\pm$ 2.5} & 14.9  {$\pm$ 3.6} &&  8.2  {$\pm$ 0.4} & 10.1  {$\pm$ 0.5} \\
\bottomrule
\end{tabular}
\caption{Model performance on BIRCO's test sets and the full original datasets}
\label{tab: test_set_whole_set_comparison}
\end{table*}

\section{Development Set}
\label{sec: appendix_dev_set}
We construct the development set by extracting triplets from unused queries and candidate pools, ensuring no overlaps between test and development set queries and passages. The dev set for DORIS-MAE has 40 queries. The dev set for Clinical-Trial has the rest 9 queries. All the other datasets have 50 queries in their dev sets. A triplet has the form of (query, relevant passage, less relevant passage). The evaluation metric in development set is the number of times a model can rank the relevant passage over the less relevant passage given the query. In particular, for Clinical-Trial, we extract all (query, score 2, score 1), (query, score 1, score 0), (query, score 2, score 0) triplets, since it uses a 3-level relevancy scale. For DORIS-MAE, we extract all (query, score x, score y) triplets such that $x-y> 0.25$, $x, y> \textrm{max score}/2$, since DORIS-MAE uses a multi-level, almost continuous relevancy scale. The number of total possible triplets is reported in Table \ref{tab: birco-beir-comparison}. In practice, for time and cost reasons, we use less than 100 triplets from each dataset for all prompt designs.

\section{Analysis of the Comparison between Datasets}
\label{sec: appendix_analysis of the comparison between birco and other datasets}
Table \ref{tab: appendix_comparison_between_birco_and_beir} presents the performance of the E5-v2-L model on BIRCO and BEIR datasets with respect to R@5, R@20, nDCG@10, MRR@10, and MAP. The maximum possible Recall@k (i.e. theoretical upper-bound) is inversely proportional to the number of relevant passages per query (R/D) if that number exceeds k. Therefore, for the R@k metric to reach approximately 100\%, the R/D for that dataset should not exceed k. Consequently, for BIRCO, R@20 can reach a maximum average of 94.3\% and is suitable for coarse-grained comparison with other datasets. In Table \ref{tab: nq_msmarco_beirs}, we compare BIRCO with NQ with respect to R@20, as the maximum R@20 for both benchmarks is near 100\%. From Table \ref{tab: nq_msmarco_beirs}, for example, we see SimLM embedding model's performance on NQ is 81.9\% R@20, but only reaches 49.6\% R@20, even though BIRCO's max possible R@20 is 94\%, this is still conclusive evidence to indicate BIRCO is a much harder benchmark than NQ, at least w.r.t SimLM. 

For nDCG@10, the maximum value can always reach 100\%. However, in a dataset where many passages per query have a non-zero relevance score, such as DORIS-MAE, which uses a fine-grained relevance scale, or TREC-COVID, where many passages are relevant, the nDCG@10 value would be inflated. This is why the random baseline on DORIS-MAE can achieve 49.9\% nDCG@10.

Furthermore, a smaller candidate pool size (or corpus size if all queries share the same candidate pool) would result in non-increasing task difficulty and non-decreasing model performance.

We therefore conclude that the average nDCG@10 metric reported on BIRCO is slightly elevated due to its smaller candidate pool size and its inclusion of DORIS-MAE and Clinical-Trial. Similarly, the average nDCG@10 reported on BEIR is slightly elevated due to several BEIR datasets having large R/D values. 
E5-v2 achieves an average nDCG of 38.6\% on BIRCO, which is significantly lower than BEIR's average 50.0\% nDCG.

Table \ref{tab: appendix_comparison_between_birco_and_beir} also allows for the comparison of the nDCG@10 of any two datasets with the same R/D value. For instance, RELIC only has 11.2\% nDCG@10, suggesting it is the most challenging task with respect to E5-v2, despite its smaller candidate pool size.

\section{Definition of Facets in BIRCO Queries}
\label{sec: appendix_procedure to calculate facets from queries}
This section elaborates on the definition of facets for each BIRCO query.

The DORIS-MAE dataset has a hierarchical, aspect-based structure. This structure divides each query into a two-level tree of semantic components, or facets. Consequently, we measure the degree of multi-facetedness by counting the number of facets and sub-facets present in each query. For ArguAna, we use GPT4 to count the number of unique arguments and their corresponding evidence within a query. For WhatsThatBook, we use GPT4 to count the number of unique details about a book provided in the user query. For Clinical-Trial, we use GPT4 to count the number of unique symptoms reported in a patient's case. For RELIC, given that the [masked sentence] has both a preceding and a subsequent context, we use GPT4 to determine if these two contexts convey the same idea. If they do not, they are counted as two separate facets. To ensure the accuracy of GPT4's performance, we manually inspect its output. For full results, see Table \ref{tab: birco-multifacetedness}. For GPT4 prompts to calculate facets, see Appendix \ref{ssec: appendix_prompts_for_multifacetedness}. 

\section{Scoring-based Framework Illustration}
\label{sec: appendix_framework_illustration}

For clarity sake, see Figure \ref{fig: SER_DeAGO_Framework} for an illustration of our proposed framework and three distinct pipelines: Score+O (top), Reason+O (middle), Subtask+O (bottom). See Figure \ref{fig: example_decision_criteria} for an example set of decision criteria produced by Reason+O.

Note, for each query, we select a single set of decision criteria which are used across all candidate passages. An alternative, not investigated in this paper, is to condition the decision criteria on both the query and the passage, increasing the specificity of the criteria but also increasing their variability. 

\section{Error Analysis for Models}
\label{sec: appendix_error analysis for models}

From manual inspection, we observed a few sources of common errors made by LLMs. In the Reason+O pipeline, a low-quality set of decision criteria typically led to low performance on this query. 

From Table \ref{tab: model_performance_combined_app}, we see Subtask+O\textsubscript{\scriptsize{GPT}} obtains better or comparable performance compared to Reason+O\textsubscript{\scriptsize{GPT}}, indicating reducing task complexity and decomposing task into sub-tasks offer more stability for GPT4. In particular, we see significant improvements of Subtask+O\textsubscript{\scriptsize{GPT}} over Reason+O\textsubscript{\scriptsize{GPT}} on the ArguAna dataset, showing that for certain complex tasks, a well-decomposed task structure enables more accurate scoring and annotation.

\subsection{Comparison vs Scoring}

Both comparison-based approaches and scoring-based approaches provide improvements in performance compared to previous models. The comparison-based approach does not provide interpretable reasoning, but still achieves comparable performances on DORIS-MAE and WhatsThatBook, and even slightly better performances on RELIC, suggesting the potential value of exposing multiple passages to LLM for comparison at the same time. On the other hand, scoring-based approaches that annotate the decomposed sub-tasks offer more stability and natural language reasoning across datasets.

\subsection{Features that impact Task Difficulty}
We examine several features of the datasets that may be contributing to the difficulty of retrieval, with a focus on GPT4 with direct scoring. We investigate various statistical features of the query to determine their potential impacts on task difficulty. For each query in each dataset, we use the average nDCG@10 of the query evaluated on Score+O\textsubscript{GPT4}, Reason+O\textsubscript{GPT4}, and Subtask+O\textsubscript{GPT4} as an indicator of the task difficulty in relation to this query. 

A higher average nDCG indicates that all three methods retrieved relevant passages for the query, while a lower nDCG suggests these methods failed to do so. The features we investigated include: query length, number of facets in a query, the query's lexical overlap with its relevant passage(s), the query's lexical overlap with all passages in the candidate pool, and the size of the candidate pool. 

We fit a linear regression for each feature and dataset. We found the following statistically significant results: a positive linear relationship between the lexical overlap of a query and its relevant passage(s), and the nDCG scores for ArguAna (p<0.005) and WhatsThatBook (p<0.0005); a negative linear relationship between the lexical overlap of a query and all passages, and the NDCG scores for RELIC (p<0.0005). Note, even for these statistically significant features, less than 15\% variance could be explained by these features, suggesting the complex nature of predicting task difficulty. 

\section{Decontamination Filtering Procedure for BIRCO}
\label{sec: appendix_decontamination}

The DORIS-MAE task requires identifying citations for specific computer science papers. GPT4 was unable to do this successfully for any of the queries. Similarly for ArguAna, the GPT4 generated counterargument does not indicate it has seen or recalled the correct counterargument. For both cases, human annotation is performed to determine whether contamination occurred. First, GPT4 is prompted to respond to the DORIS-MAE query with a paper abstract, and to respond to the ArguAna query with a counterargument. Then, annotators analyze whether GPT4's output paper abstract and counterargument are semantically similar to the actual relevant abstract(s) and counterargument. 
Annotators did not find any such feature to suggest GPT4's external knowledge can guide it to select relevant passages for queries in DORIS-MAE and ArguAna. 

For WhatsThatBook, we prompted GPT4 to generate its best guesses for the forgotten book title based on the user query, and if any of GPT4's generated titles share significant textual similarity with the correct book title, the query is removed. 38 queries out of 437 are removed. We also remove any passage (i.e. book description) from the corpus if GPT4 can guess the book title from the book description, suggesting it knows information beyond the book description that could potentially help it answer the query. 4817 out of 8988 book descriptions are removed. Finally, we also remove a query if its corresponding correct book description is removed, resulting in the removal of an additional 238 queries. 

For Clinical-Trial, we prompted GPT4 to generate the most relevant clinical trial for the patient based on the patient's case report. Human annotation is conducted to compare GPT4's answer with one of the relevant clinical studies. In cases where there are multiple relevant passages, we choose the relevant passage with the highest cosine similarity with GPT4's result, calculated by ada-002 embeddings. This is done to simplify and stabilize the human annotation process. In 4 queries out of 59, we find knowledge similarity between GPT4's answer and the correct answer, i.e. they mention similar treatment options. For example, GPT4 seems to have prior knowledge that thyroid hormone replacement therapy is recommended to treat hypothyroidism. These 4 queries are removed from the test set.

For RELIC, we prompted GPT4 to generate sentences that could be the masked sentence based on the query which only contains the surrounding context of the masked sentence. If any of the generated sentences share significant textual similarity (i.e. a 5-gram overlap), we remove the query. 311 out of 3823 queries are removed.

We provide all prompts for decontamination in Appendix \ref{ssec: appendix_prompts_for_decontamination}

\section{Dataset License}
\label{sec: appendix_dataset_license_appendix}

\begin{itemize}
    \item DORIS-MAE: Provided under CC-BY-NC 4.0 license.
    \item ArguAna: Provided under CC-BY-SA 4.0 license.
    \item WhatsThatBook: Provided under ODC-BY license.
    \item Clinical-Trial: Provided under CC-BY-SA 4.0 license.
    \item RELIC: Provided under MIT license.
\end{itemize}

\section{Example (Query, Passage) Pairs from the Dataset}
\label{sec: appendix_example}

\begin{figure*}[ht]
    \centering
        \includegraphics[width=\linewidth]{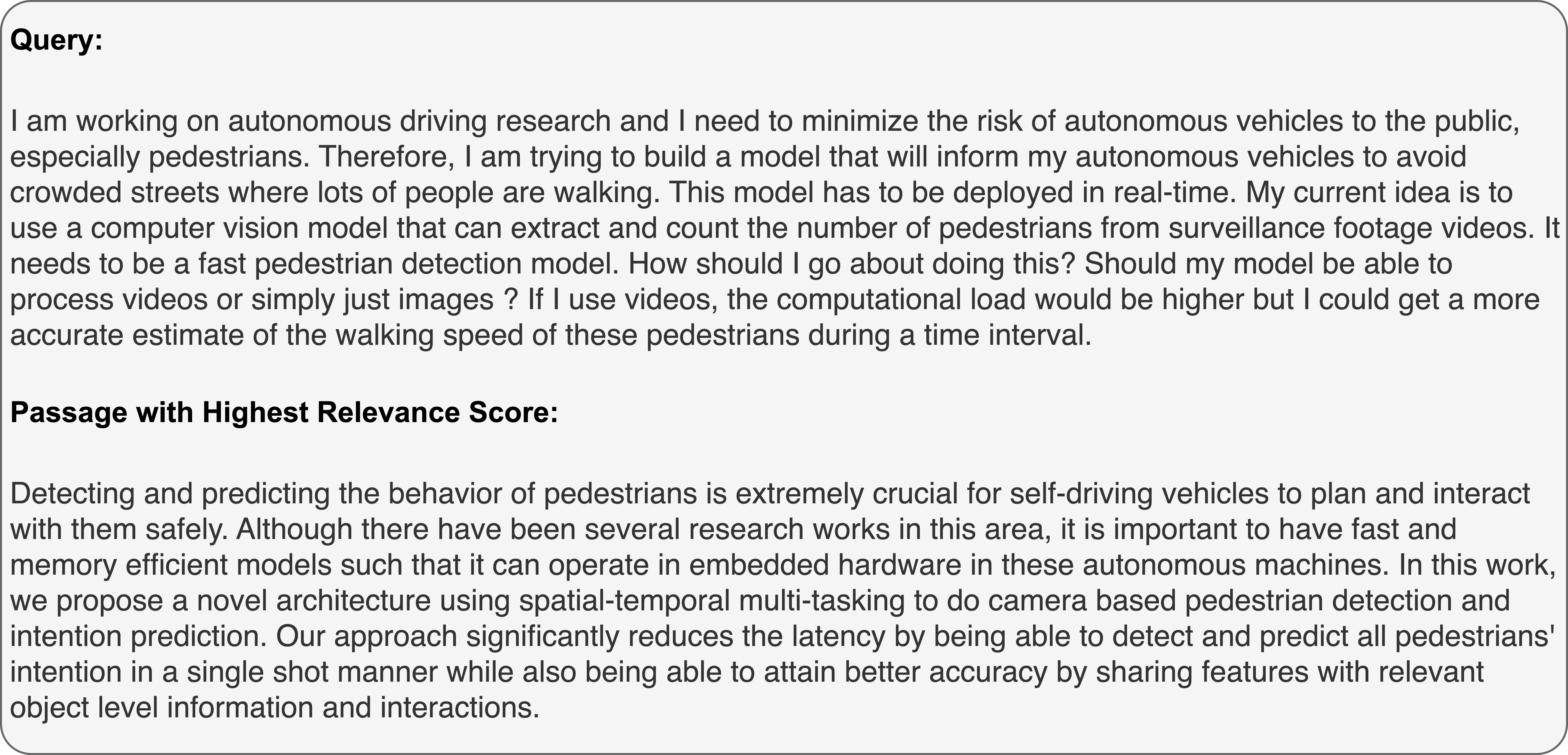}
    \caption{Example query and passage pair from DORIS-MAE}
    \label{fig: example_dm}
\end{figure*}

\begin{figure*}[ht]
    \centering
        \includegraphics[width=\linewidth]{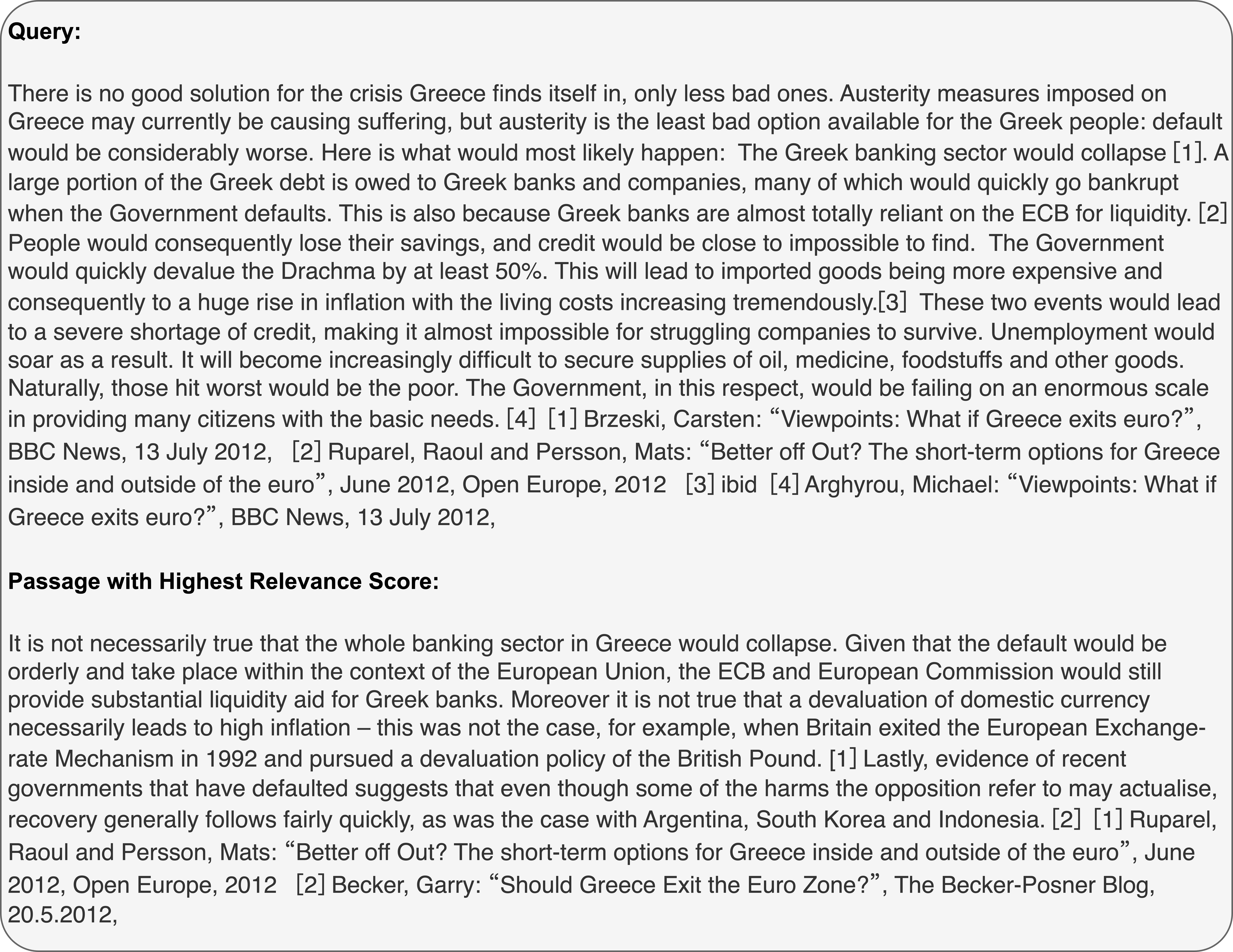}
    \caption{Example query and passage pair from ArguAna}
    \label{fig: example_arguana}
\end{figure*}

\begin{figure*}[ht]
    \centering
        \includegraphics[width=\linewidth]{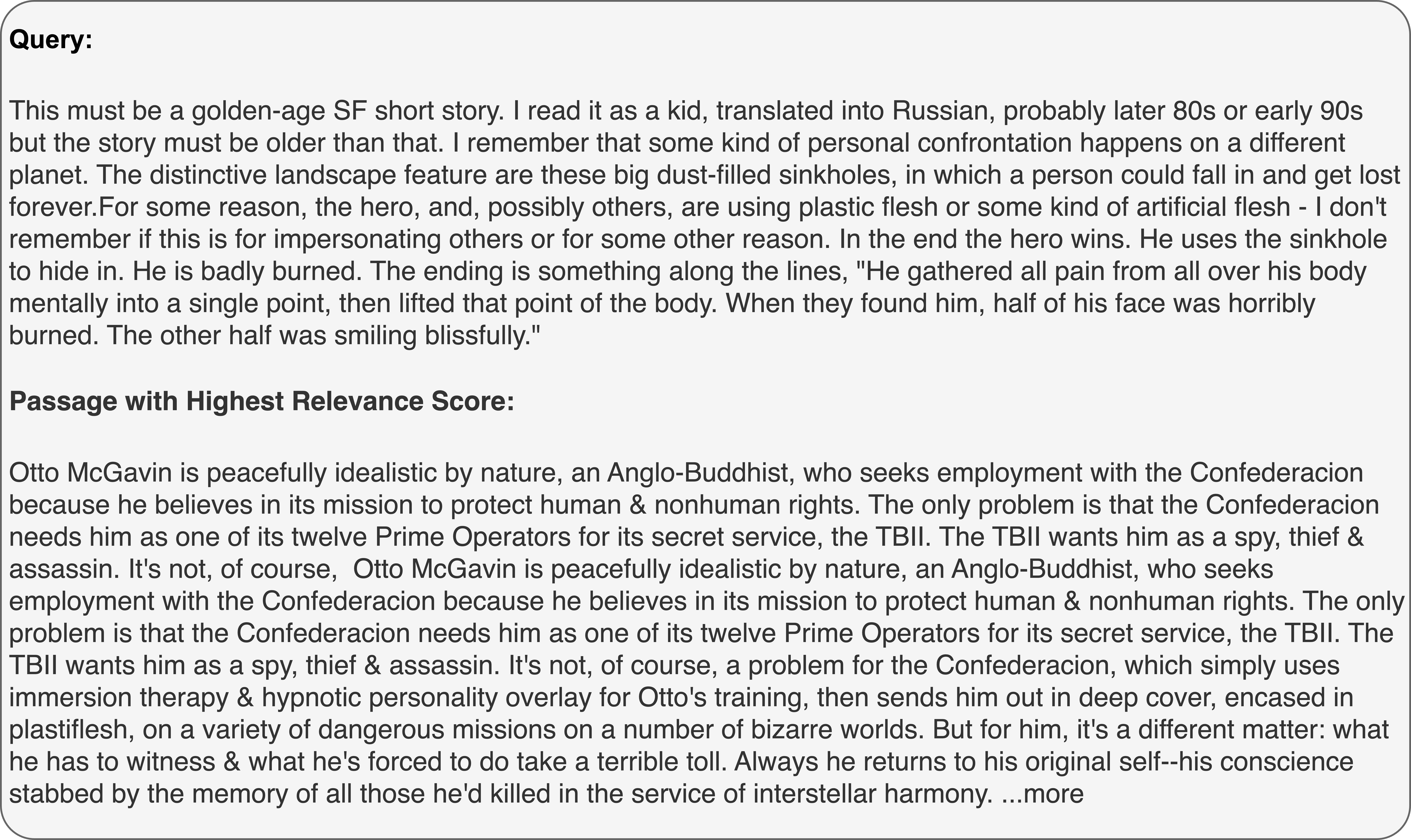}
    \caption{Example query and passage pair from WhatsThatBook}
    \label{fig: example_wtb}
\end{figure*}

\begin{figure*}[ht]
    \centering
        \includegraphics[width=\linewidth]{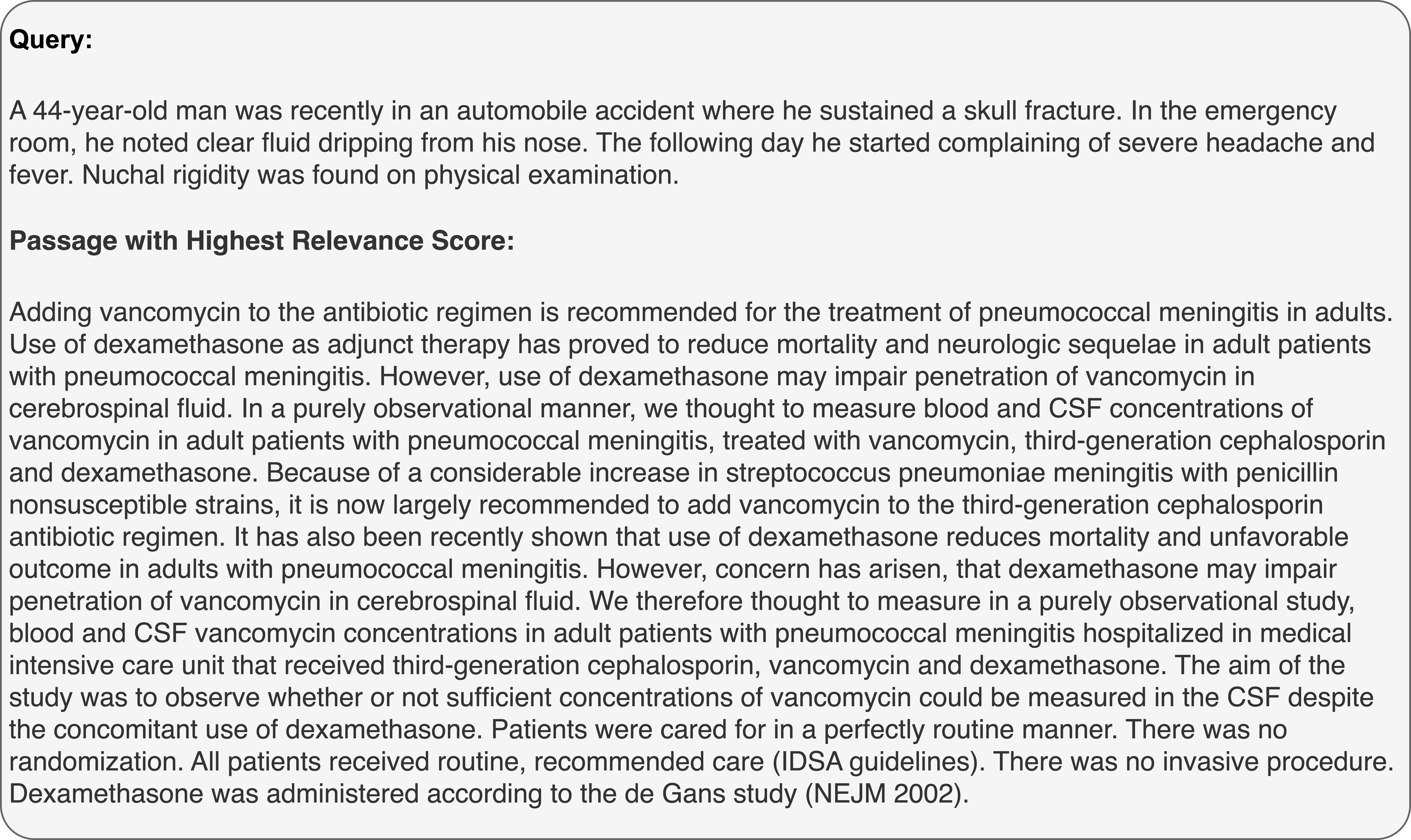}
    \caption{Example query and passage pair from Clinical-Trial}
    \label{fig: example_clinical}
\end{figure*}

\begin{figure*}[ht]
    \centering
        \includegraphics[width=\linewidth]{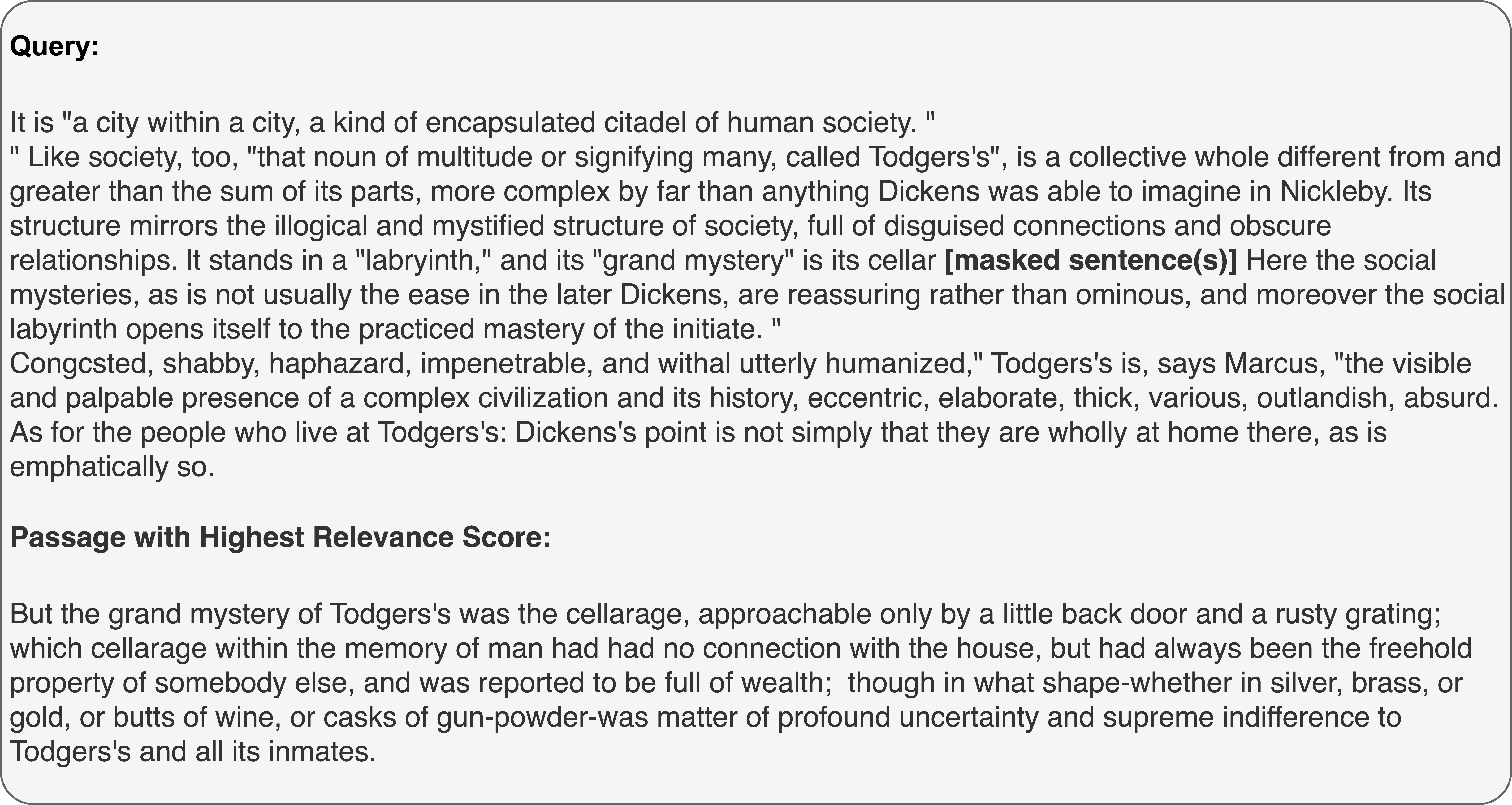}
    \caption{Example query and passage pair from RELIC}
    \label{fig: example_relic}
\end{figure*}

Please refer to Figure \ref{fig: example_dm}, \ref{fig: example_arguana}, \ref{fig: example_wtb}, \ref{fig: example_clinical}, \ref{fig: example_relic} for examples of query and passage pairs from the datasets.

\section{All Prompting Strategies}
\label{sec: appendix_prompt}
This section includes all prompting templates used in the paper.

\subsection{Prompts for Score+O}
\label{ssec: appendix_prompts_for_scoreO}
\begin{figure*}[ht]
    \centering
        \includegraphics[width=\linewidth]{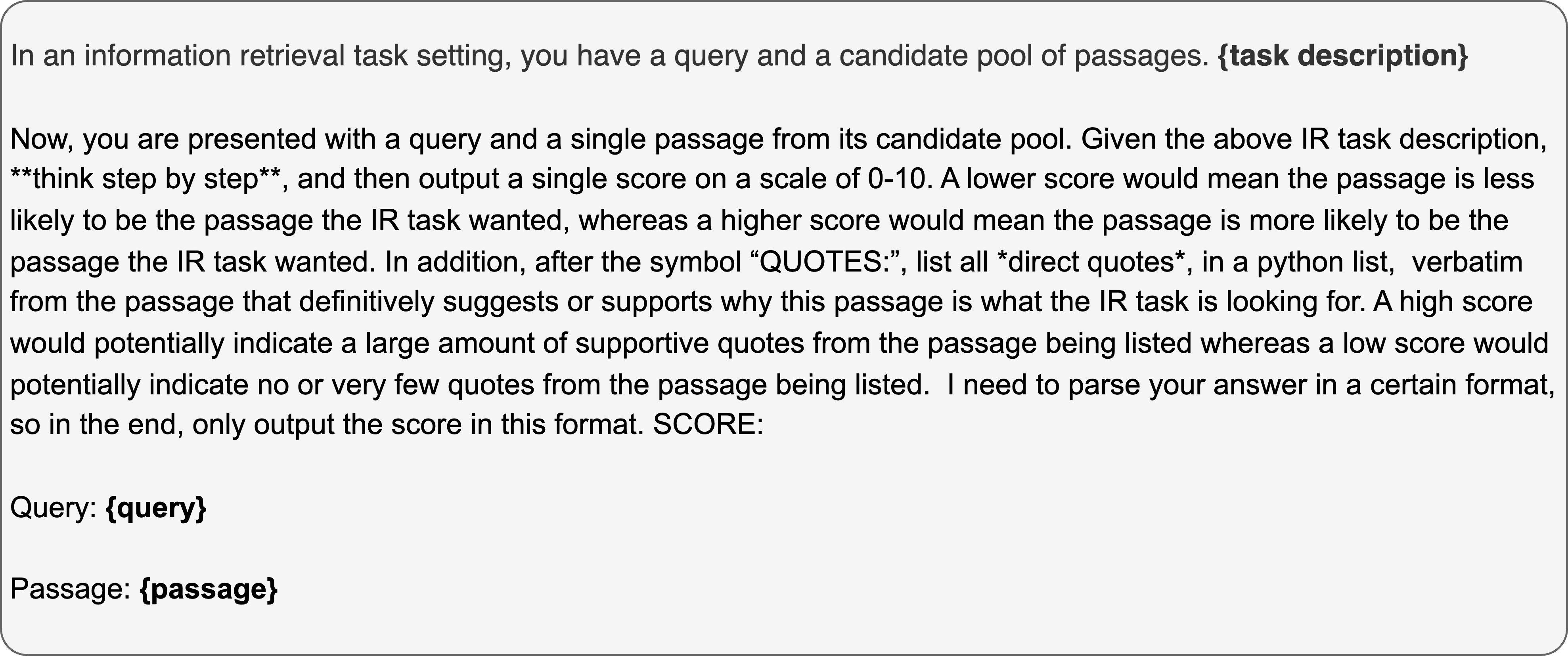}
    \caption{Prompt for Score+O}
    \label{fig: score_prompt}
\end{figure*}

Please refer to Figure \ref{fig: score_prompt} for the prompt for Score+O.

\subsection{Prompts for Reason+O}
\label{ssec: appendix_prompts_for_reasonO}
\begin{figure*}[ht]
    \centering
        \includegraphics[width=\linewidth]{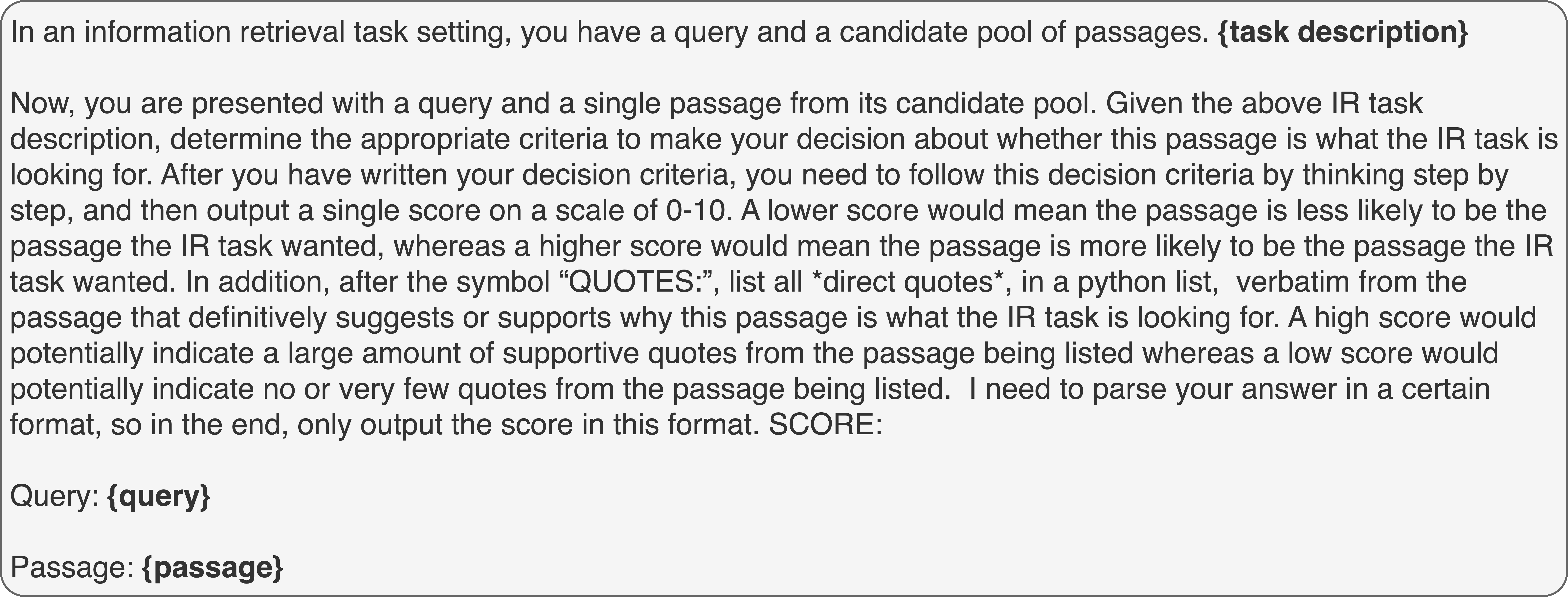}
    \caption{Prompt for Reason+O}
    \label{fig: reason_prompt}
\end{figure*}

Please refer to Figure \ref{fig: reason_prompt} for the prompt for Reason+O.

\subsection{Prompts for Score+O}
\begin{figure*}[ht]
    \centering
        \includegraphics[width=\linewidth]{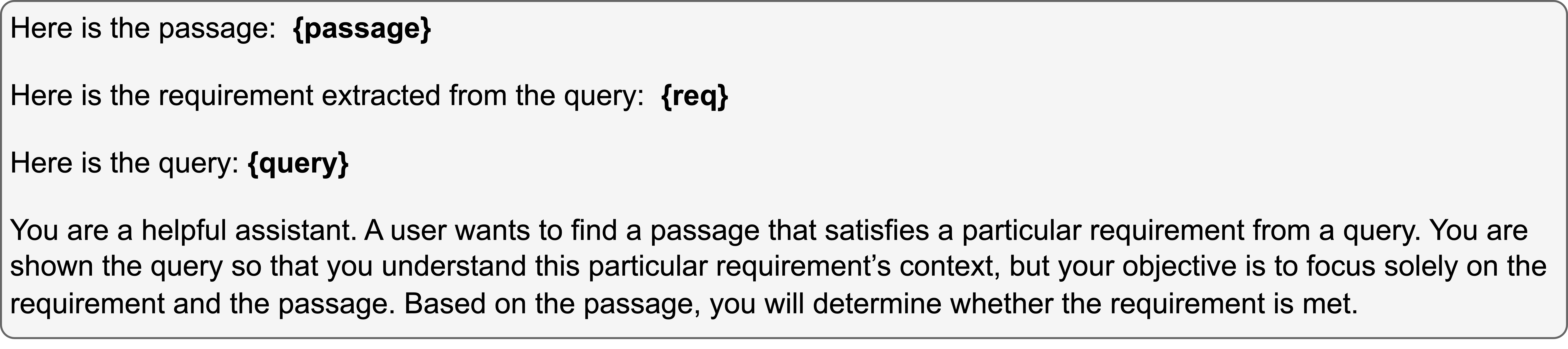}
    \caption{Prompt for Subtask+O}
    \label{fig: breakdown_prompt}
\end{figure*}

 Please refer to Figure \ref{fig: breakdown_prompt} for the prompt for Subtask+O.

\subsection{Prompts for Checking Multifacetedness}
\label{ssec: appendix_prompts_for_multifacetedness}
\begin{figure*}[ht]
    \centering
        \includegraphics[width=\linewidth]{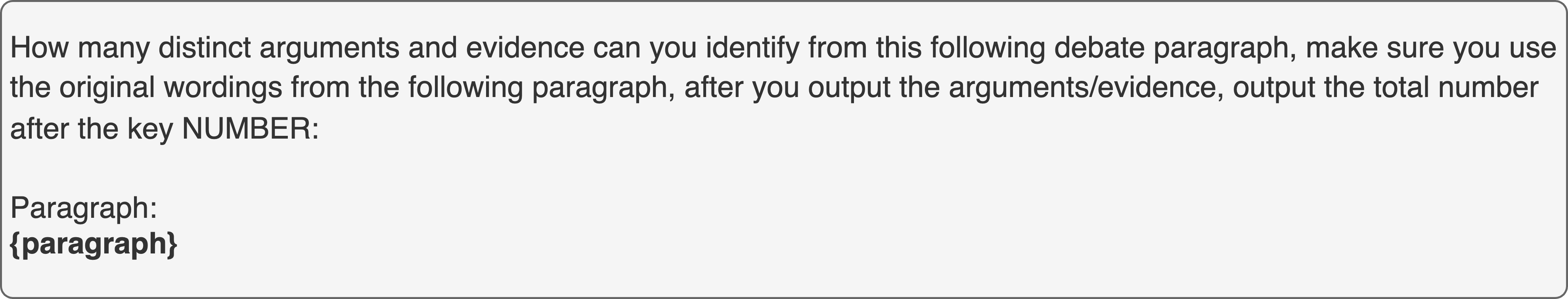}
    \caption{Prompt for getting facets for ArguAna}
    \label{fig: arguana_facet_prompt}
\end{figure*}

\begin{figure*}[ht]
    \centering
        \includegraphics[width=\linewidth]{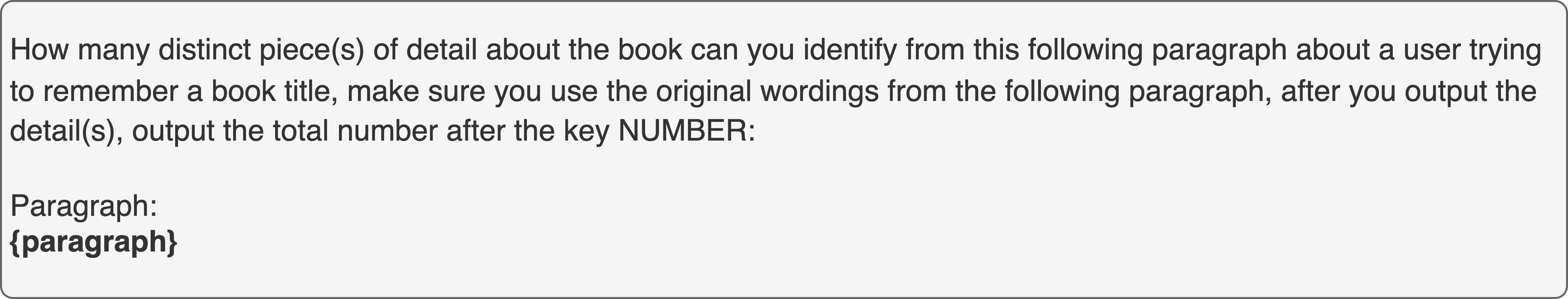}
    \caption{Prompt for getting facets for WhatsThatBook}
    \label{fig: wtb_facet_prompt}
\end{figure*}

\begin{figure*}[ht]
    \centering
        \includegraphics[width=\linewidth]{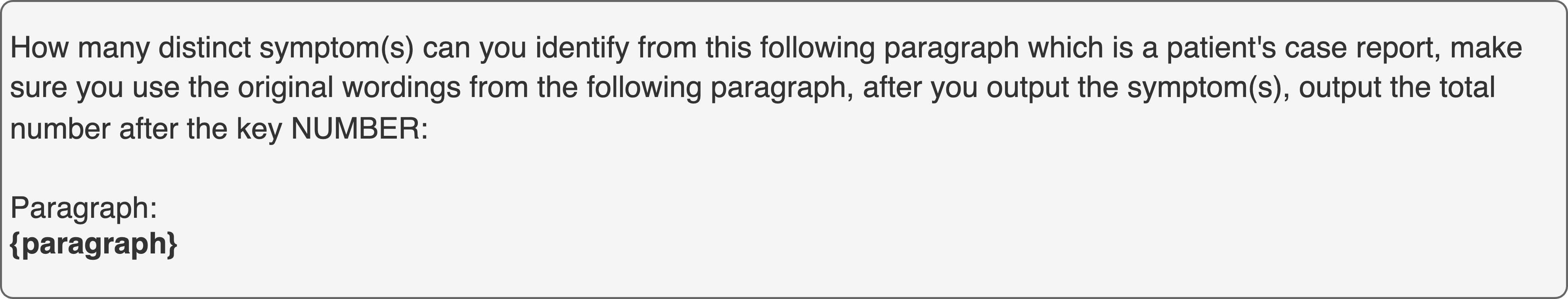}
    \caption{Prompt for getting facets for Clinical-trial}
    \label{fig: clinical-trial_facet_prompt}
\end{figure*}

\begin{figure*}[ht]
    \centering
        \includegraphics[width=\linewidth]{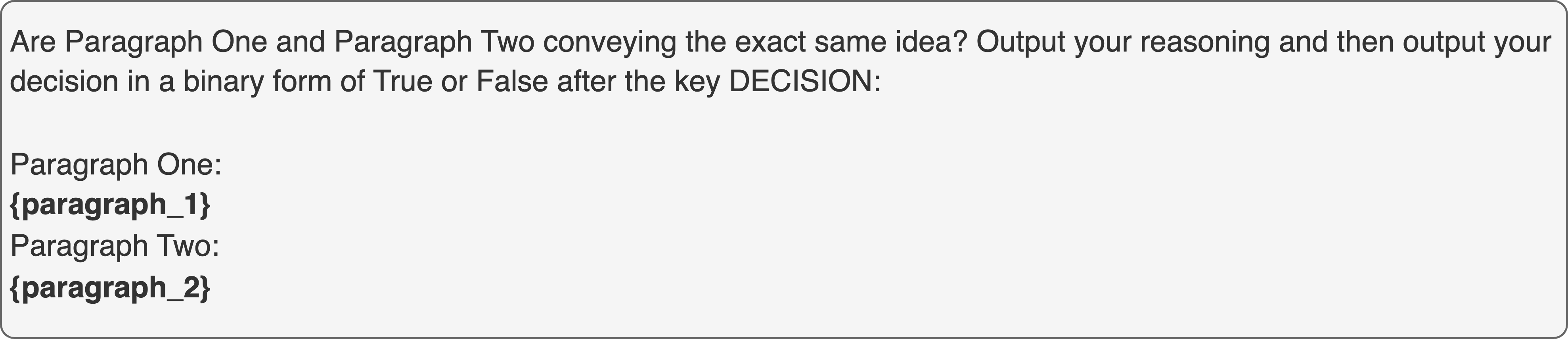}
    \caption{Prompt for getting facets for RELIC}
    \label{fig: relic_facet_prompt}
\end{figure*}

Please refer to Figure \ref{fig: arguana_facet_prompt}, \ref{fig: wtb_facet_prompt}, \ref{fig: clinical-trial_facet_prompt}, \ref{fig: relic_facet_prompt} for prompts for checking the multifacetedness in datasets.

\subsection{Prompts for Decontamination}
\label{ssec: appendix_prompts_for_decontamination}
\begin{figure*}[ht]
    \centering
        \includegraphics[width=\linewidth]{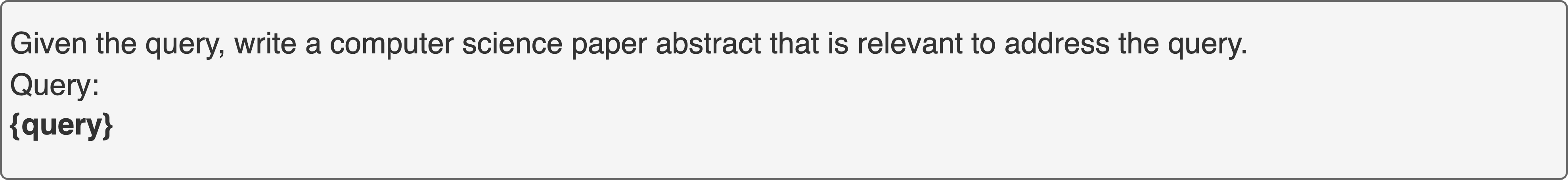}
    \caption{Prompt for DORIS-MAE decontamination}
    \label{fig: doris-mae_decontamination_prompt}
\end{figure*}

\begin{figure*}[ht]
    \centering
        \includegraphics[width=\linewidth]{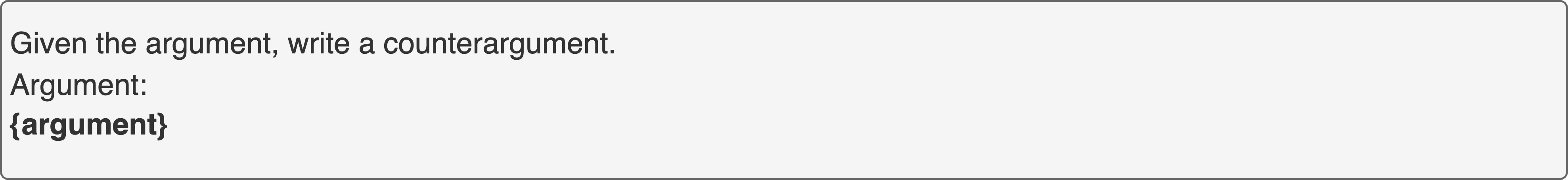}
    \caption{Prompt for ArguAna decontamination}
    \label{fig: arguana_decontamination_prompt}
\end{figure*}

\begin{figure*}[ht]
    \centering
        \includegraphics[width=\linewidth]{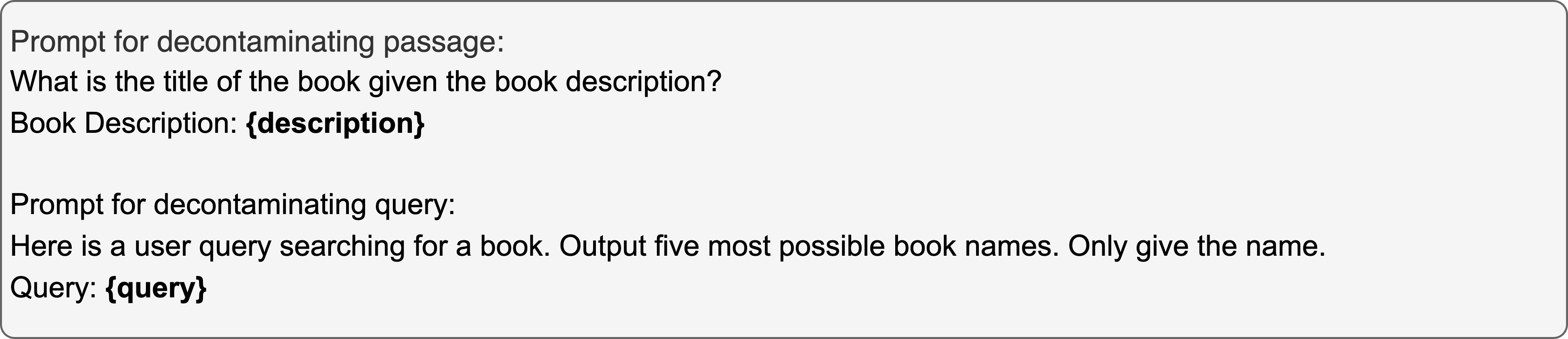}
    \caption{Prompts for WhatsThatBook decontamination}
    \label{fig: wtb_decontamination_prompt}
\end{figure*}

\begin{figure*}[ht]
    \centering
        \includegraphics[width=\linewidth]{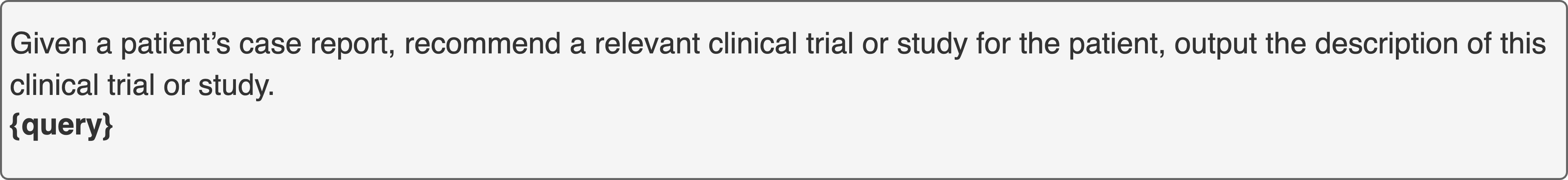}
    \caption{Prompt for Clinical-trial decontamination}
    \label{fig: clinical-trial_decontamination_prompt}
\end{figure*}

\begin{figure*}[ht]
    \centering
        \includegraphics[width=\linewidth]{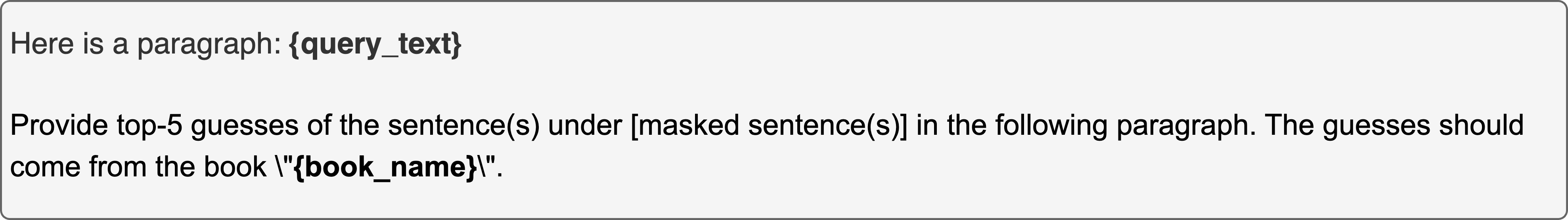}
    \caption{Prompt for RELIC decontamination}
    \label{fig: relic_decontamination_prompt}
\end{figure*}

Please refer to Figure \ref{fig: doris-mae_decontamination_prompt}, \ref{fig: arguana_decontamination_prompt}, \ref{fig: wtb_decontamination_prompt}, \ref{fig: clinical-trial_decontamination_prompt}, \ref{fig: relic_decontamination_prompt} for prompts for decontamination in BIRCO.

\begin{figure*}[ht]
    \centering
        \includegraphics[width=\linewidth]{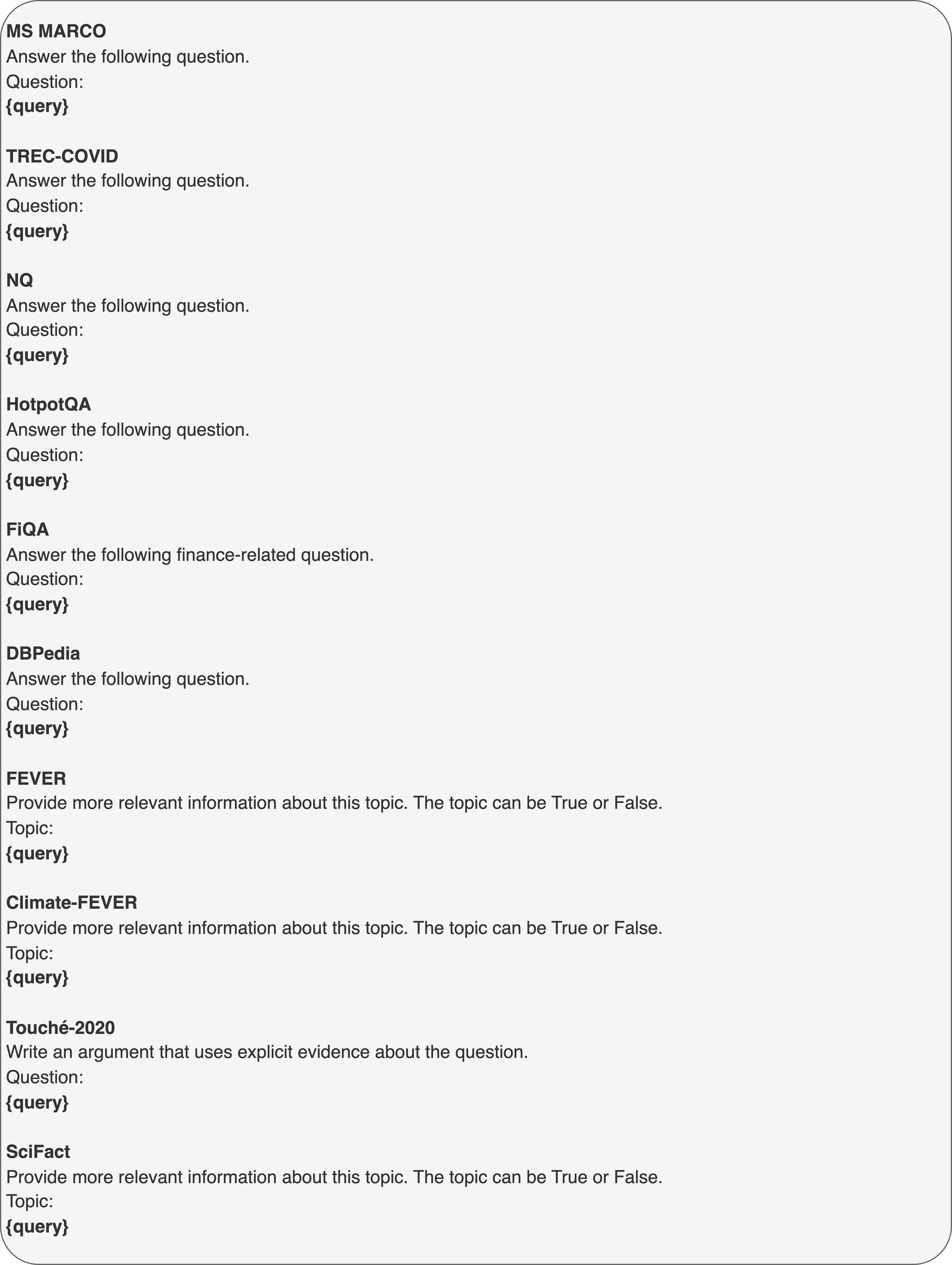}
    \caption{Prompts for BEIR decontamination}
    \label{fig: beir_decontamination_prompt}
\end{figure*}

Please refer to Figure \ref{fig: beir_decontamination_prompt} for prompts for decontamination in BEIR.

\end{document}